\documentclass[aps,prc,groupedaddress,showpacs]{revtex4}
\usepackage{graphicx}
\usepackage{amsmath,bm,calc}
\usepackage[psamsfonts]{amssymb}

\begin{document}


\title{Mean-field approach to nuclear structure
with semi-realistic nucleon-nucleon interactions}


\author{H. Nakada}
\email[E-mail:\,\,]{nakada@faculty.chiba-u.jp}
\affiliation{Department of Physics, Graduate School of Science,
 Chiba University\\
Yayoi-cho 1-33, Inage, Chiba 263-8522, Japan}


\date{\today}

\begin{abstract}
Semi-realistic nucleon-nucleon interactions
applicable to the self-consistent mean-field
(both Hartree-Fock and Hartree-Fock-Bogolyubov) calculations
are developed,
by modifying the M3Y interaction.
The modification is made so as to reproduce
binding energies and rms matter radii of doubly magic nuclei,
single-particle levels in $^{208}$Pb,
and even-odd mass differences of the Sn isotopes.
We find parameter-sets with and without the tensor force.
The new interactions are further checked by
the saturation properties of the uniform nuclear matter
including the Landau-Migdal parameters.
By the mean-field calculations,
interaction-dependence of the neutron drip line
is investigated for the O, Ca and Ni isotopes,
and of the single-particle energies
for the $N=16$, $32$, $50$, $82$ and $Z=50$ nuclei.
Results of the semi-realistic interactions
including the tensor force are in fair agreement
with available experimental data for all of these properties.
\end{abstract}


\maketitle



\section{Introduction\label{sec:intro}}

Mean-field (MF) theories provide us with a good first approximation
to the nuclear structure problems.
They are able to describe the saturation and the shell structure
simultaneously, both of which are basic nuclear properties,
based on effective nucleon-nucleon ($NN$) interactions.
As far as we constrain to the non-relativistic approaches,
most of the MF calculations have been performed
with the Skyrme interaction~\cite{ref:VB72}.
Finite-range interactions have rarely been applied,
except the Gogny interaction~\cite{ref:Gogny}
which has the Gaussian form for the central channels.
Most popular parameter-sets of the Skyrme and the Gogny interactions
have been adjusted mainly to the nuclear properties
around the $\beta$-stability.
However, it is a question whether
such phenomenological effective interactions
work well for nuclei far off the $\beta$-stability.
For instance, whereas role of the tensor force
in nucleus-dependence of the MF
has attracted interest~\cite{ref:Vtn,ref:LBB07},
the tensor force is usually ignored
in those parameter-sets.

Although direct application of the bare $NN$ (and $NNN$) interaction
to the nuclear structure problems~\cite{ref:micro,ref:NCSM,ref:HPDH08}
is yet limited to light nuclei or made to medium-mass nuclei
but with limited accuracy,
guide from microscopic theories
will be valuable even in heavy-mass nuclei.
The Michigan 3-range Yukawa (M3Y) interaction~\cite{ref:M3Y},
which was derived by fitting the Yukawa functions
to Brueckner's $G$-matrix,
has been used in nuclear structure
as well as in low-energy nuclear reaction studies.
There have been a few attempts applying the M3Y-type interaction
to MF calculations~\cite{ref:HL98,ref:Nak03}.
In Ref.~\cite{ref:Nak03},
the author has developed an M3Y-type interaction
which is applicable to the Hartree-Fock (HF) calculations.
The original M3Y interaction is incapable of reproducing the saturation
and the spin-orbit splitting within the MF regime.
To cure this problem a density-dependent contact term has been added
and some of the strength parameters have been modified.
Such interactions,
which have originally been derived from microscopic theories
but are slightly modified from phenomenological standpoints,
may be called \textit{semi-realistic} interactions.
It has been shown~\cite{ref:Nak03,ref:Nak04}
that semi-realistic $NN$ interactions could give different shell structure
from the widely used Skyrme and Gogny interactions.
However, the pairing properties have not been taken into account
in the parameter-set M3Y-P2 that was proposed in Ref.~\cite{ref:Nak03}.
This implies that the singlet-even channel in M3Y-P2
is not quite appropriate
as long as the pairing interaction is taken to be consistent
with the HF interaction,
while this problem seems to be masked in the HF approximation.
In order to apply semi-realistic interactions
to the MF studies extensively,
we explore new parameter-sets of the M3Y-type interaction,
taking the pairing properties into account.
Special attention is paid also to role of the tensor force.
Using the recently developed algorithm~\cite{ref:NS02,ref:Nak06,ref:Nak08},
we apply the new semi-realistic interactions
to the Hartree-Fock-Bogolyubov (HFB) as well as to the HF calculations
of medium- to heavy-mass spherical nuclei.

\section{M3Y-type interaction\label{sec:M3Y}}

We express a non-relativistic nuclear effective Hamiltonian by
\begin{equation}
H_N = K + V_N\,;\quad K = \sum_i \frac{\mathbf{p}_i^2}{2M}\,,\quad
V_N = \sum_{i<j} v_{ij}\,,
\label{eq:H_N}\end{equation}
with $i$ and $j$ representing the indices of individual nucleons.
For the effective $NN$ interaction $v_{ij}$,
we consider the following form,
\begin{eqnarray} v_{ij} &=& v_{ij}^{(\mathrm{C})}
 + v_{ij}^{(\mathrm{LS})} + v_{ij}^{(\mathrm{TN})}
 + v_{ij}^{(\mathrm{DD})}\,;\nonumber\\
v_{ij}^{(\mathrm{C})} &=& \sum_n \big(t_n^{(\mathrm{SE})} P_\mathrm{SE}
+ t_n^{(\mathrm{TE})} P_\mathrm{TE} + t_n^{(\mathrm{SO})} P_\mathrm{SO}
+ t_n^{(\mathrm{TO})} P_\mathrm{TO}\big)
 f_n^{(\mathrm{C})} (r_{ij})\,,\nonumber\\
v_{ij}^{(\mathrm{LS})} &=& \sum_n \big(t_n^{(\mathrm{LSE})} P_\mathrm{TE}
 + t_n^{(\mathrm{LSO})} P_\mathrm{TO}\big)
 f_n^{(\mathrm{LS})} (r_{ij})\,\mathbf{L}_{ij}\cdot
(\mathbf{s}_i+\mathbf{s}_j)\,,\nonumber\\
v_{ij}^{(\mathrm{TN})} &=& \sum_n \big(t_n^{(\mathrm{TNE})} P_\mathrm{TE}
 + t_n^{(\mathrm{TNO})} P_\mathrm{TO}\big)
 f_n^{(\mathrm{TN})} (r_{ij})\, r_{ij}^2 S_{ij}\,,\nonumber\\
v_{ij}^{(\mathrm{DD})} &=& \big(t_\rho^{(\mathrm{SE})} P_\mathrm{SE}\cdot
 [\rho(\mathbf{r}_i)]^{\alpha^{(\mathrm{SE})}}
 + t_\rho^{(\mathrm{TE})} P_\mathrm{TE}\cdot
 [\rho(\mathbf{r}_i)]^{\alpha^{(\mathrm{TE})}}\big)
 \,\delta(\mathbf{r}_{ij})\,,
\label{eq:effint}\end{eqnarray}
where $\mathbf{r}_{ij}= \mathbf{r}_i - \mathbf{r}_j$,
$r_{ij}=|\mathbf{r}_{ij}|$,
$\mathbf{p}_{ij}= (\mathbf{p}_i - \mathbf{p}_j)/2$,
$\mathbf{L}_{ij}= \mathbf{r}_{ij}\times \mathbf{p}_{ij}$,
$S_{ij}= 4\,[3(\mathbf{s}_i\cdot\hat{\mathbf{r}}_{ij})
(\mathbf{s}_j\cdot\hat{\mathbf{r}}_{ij})
- \mathbf{s}_i\cdot\mathbf{s}_j ]$
with $\hat{\mathbf{r}}_{ij}=\mathbf{r}_{ij}/r_{ij}$,
and $\rho(\mathbf{r})$ denotes the nucleon density.
The Yukawa function $f_n(r)=e^{-\mu_n r}/\mu_n r$
is assumed for all channels except $v^{(\mathrm{DD})}$
in the M3Y-type interactions.
The projection operators
on the singlet-even (SE), triplet-even (TE), singlet-odd (SO)
and triplet-odd (TO) two-particle states are defined as
\begin{eqnarray}
P_\mathrm{SE} = \frac{1-P_\sigma}{2}\,\frac{1+P_\tau}{2}\,,
\quad P_\mathrm{TE} = \frac{1+P_\sigma}{2}\,\frac{1-P_\tau}{2}\,,
\nonumber\\
P_\mathrm{SO} = \frac{1-P_\sigma}{2}\,\frac{1-P_\tau}{2}\,.
\quad P_\mathrm{TO} = \frac{1+P_\sigma}{2}\,\frac{1+P_\tau}{2}\,,
\label{eq:proj_T}\end{eqnarray}
where $P_\sigma$ ($P_\tau$) expresses
the spin (isospin) exchange operator.

We shall start from the M3Y-Paris interaction~\cite{ref:M3Y-P},
which will be denoted by M3Y-P0 in this article
as in Ref.~\cite{ref:Nak03}.
We change none of the range parameters $\mu_n$ of M3Y-P0
in $v^{(\mathrm{C})}$, $v^{(\mathrm{LS})}$ and $v^{(\mathrm{TN})}$.
In M3Y-P0, the longest range part in $v^{(\mathrm{C})}$
is kept identical to the central channels
of the one-pion exchange potential (OPEP), $v^{(\mathrm{C})}_\mathrm{OPEP}$.
We also maintain this reasonable assumption.
As is well known, the spin-orbit ($\ell s$) splitting plays
a significant role in the nuclear shell structure.
Even though higher-order effects may account for
the observed $\ell s$ splitting~\cite{ref:LS},
it is desired to enhance $v^{(\mathrm{LS})}$ in order to describe
the shell structure within the MF regime.
We here use an overall enhancement factor to $v^\mathrm{(LS)}$,
which is determined from the single-particle (s.p.) spectrum
of $^{208}$Pb,
as will be shown in Sec.~\ref{sec:DMprop}.
Influence of the tensor force on s.p. energies
is a current topic,
which could be relevant to the new magic numbers
in unstable nuclei~\cite{ref:Vtn}.
We develop two parameter-sets having $v^{(\mathrm{TN})}$
without any modification from M3Y-P0,
as well as a parameter-set in which we impose $v^{(\mathrm{TN})}=0$.

The saturation properties are important to describe many nuclei
in a wide mass range.
Since it is still hard to describe accurately the saturation properties
by the bare $NN$ (and $NNN$) interaction
despite certain progress~\cite{ref:micro},
it will be appropriate to modify realistic effective interaction
so as to reproduce the saturation properties.
Density-dependence in the effective interaction has been known
to be essential in obtaining the saturation.
We therefore add a density-dependent contact force
$v^{(\mathrm{DD})}$~\cite{ref:Nak03}.
The parameter $\alpha^{(\mathrm{TE})}$ in $v^{(\mathrm{DD})}$,
power to $\rho$, is taken to be $1/3$,
by which the incompressibility $\mathcal{K}$ becomes
close to a reasonable value as shown later.
On the other hand, $\alpha^{(\mathrm{SE})}$ is not
quite sensitive to $\mathcal{K}$,
because the major source of the saturation lies in the TE channel,
not in the SE channel~\cite{ref:Bet71}.
Although we simply assumed
$\alpha^{(\mathrm{SE})}=\alpha^{(\mathrm{TE})}=1/3$
in M3Y-P2~\cite{ref:Nak03},
this assumption makes it difficult to reproduce pairing properties
and to avoid instability of the neutron matter~\cite{ref:NS02,ref:Nak03}
simultaneously.
To overcome this problem,
we adopt $\alpha^{(\mathrm{SE})}=1$ in the new parameter-sets.
The difference between $\alpha^{(\mathrm{SE})}$
and $\alpha^{(\mathrm{TE})}$ may be attributed
to the difference in origin of the $\rho$-dependence;
the short-range repulsion in the bare $NN$ interaction
in the SE channel
while primarily the tensor force in the TE channel.
Validity of the choice $\alpha^{(\mathrm{SE})}=1$
is further discussed in Sec.~\ref{sec:NMprop}.

The remaining parameters are $t_n$ in $v^{(\mathrm{C})}$
(except those of $v^{(\mathrm{C})}_\mathrm{OPEP}$)
and $t_\rho$ in $v^{(\mathrm{DD})}$.
We fit them to the measured binding energies of $^{16}$O and $^{208}$Pb,
in the HF approximation (see Sec.~\ref{sec:DMprop}).
The proton and neutron Fermi energies of $^{208}$Pb,
which are primarily relevant to the symmetry energy,
are checked additionally.
To determine $t_n^{(\mathrm{SE})}$ and $t_\rho^{(\mathrm{SE})}$,
we also use the even-odd mass differences of the Sn isotopes,
by comparing results of the Hartree-Fock-Bogolyubov (HFB) calculations
with the experimental values (see Sec.~\ref{sec:pairing}).
The new parameter-sets of the semi-realistic M3Y-type interaction,
M3Y-P3 to P5, are tabulated in Table~\ref{tab:param_M3Y}.
For comparison, M3Y-P0 and P2 are also shown.
In the set M3Y-P3, we keep both $v^{(\mathrm{TN})}$
and the odd-channel (SO and TO) strengths in $v^{(\mathrm{C})}$
of M3Y-P0.
The set M3Y-P4 is obtained by assuming $v^{(\mathrm{TN})}=0$,
while changing $t_n^{(\mathrm{SO})}$ and $t_n^{(\mathrm{TO})}$ ($n=1,2$)
substantially.
In the set M3Y-P5,
we somewhat modify $t_2^{(\mathrm{SO})}$ and $t_2^{(\mathrm{TO})}$
while keeping $v^{(\mathrm{TN})}$,
so as to reproduce the binding energies of
several doubly magic nuclei better than M3Y-P3,
as will be shown in Sec.~\ref{sec:DMprop}.
Thus the number of adjusted parameters are 7, 11 and 9
(including the overall enhancement factor to $v^{(\mathrm{LS})}$)
for M3Y-P3, P4 and P5, respectively.
It will be useful to compare results of these parameter-sets
for pinning down which part of the interaction is important
to individual physical quantities.
In particular, role of the tensor force will be of interest.
It is remarked that, while schematic tensor forces have been
introduced into some of the recent MF studies~\cite{ref:OMA06,ref:LBB07},
the present $v^{(\mathrm{TN})}$ in M3Y-P3 and P5 is much more realistic.

\begin{table}
\begin{center}
\caption{Parameters of M3Y-type interactions.
\label{tab:param_M3Y}}
\begin{tabular}{ccr@{.}lr@{.}lr@{.}lr@{.}lr@{.}l}
\hline\hline
parameters && \multicolumn{2}{c}{~~M3Y-P0~~} &
 \multicolumn{2}{c}{~~M3Y-P2~~} & \multicolumn{2}{c}{~~M3Y-P3~~} &
 \multicolumn{2}{c}{~~M3Y-P4~~} & \multicolumn{2}{c}{~~M3Y-P5~~} \\
 \hline
$1/\mu_1^{(\mathrm{C})}$ &(fm)&
 $0$&$25$ & $0$&$25$ & $0$&$25$ & $0$&$25$ & $0$&$25$ \\
$t_1^{(\mathrm{SE})}$ &(MeV)& $11466$& & $8027$&
 & $8027$& & $8027$& & $8027$& \\
$t_1^{(\mathrm{TE})}$ &(MeV)& $13967$& & $6080$&
 & $7130$& & $5503$& & $5576$& \\
$t_1^{(\mathrm{SO})}$ &(MeV)& $-1418$& & $-11900$&
 & $-1418$& & $-12000$& & $-1418$& \\
$t_1^{(\mathrm{TO})}$ &(MeV)& $11345$& & $3800$&
 & $11345$& & $3700$& & $11345$& \\
$1/\mu_2^{(\mathrm{C})}$ &(fm)&
 $0$&$40$ & $0$&$40$ & $0$&$40$ & $0$&$40$ & $0$&$40$ \\
$t_2^{(\mathrm{SE})}$ &(MeV)& $-3556$& & $-2880$&
 & $-2637$& & $-2637$& & $-2650$& \\
$t_2^{(\mathrm{TE})}$ &(MeV)& $-4594$& & $-4266$&
 & $-4594$& & $-4183$& & $-4170$& \\
$t_2^{(\mathrm{SO})}$ &(MeV)& $950$& & $2730$&
 & $950$& & $4500$& & $2880$& \\
$t_2^{(\mathrm{TO})}$ &(MeV)& $-1900$& & $-780$&
 & $-1900$& & $-1000$& & $-1780$& \\
$1/\mu_3^{(\mathrm{C})}$ &(fm)&
 $1$&$414$ & $1$&$414$ & $1$&$414$ & $1$&$414$ & $1$&$414$ \\
$t_3^{(\mathrm{SE})}$ &(MeV)& $-10$&$463$ & $-10$&$463$ & $-10$&$463$
 & $-10$&$463$ & $-10$&$463$ \\
$t_3^{(\mathrm{TE})}$ &(MeV)& $-10$&$463$ & $-10$&$463$ & $-10$&$463$
 & $-10$&$463$ & $-10$&$463$ \\
$t_3^{(\mathrm{SO})}$ &(MeV)& $31$&$389$ & $31$&$389$ & $31$&$389$
 & $31$&$389$ & $31$&$389$ \\
$t_3^{(\mathrm{TO})}$ &(MeV)& $3$&$488$ & $3$&$488$ & $3$&$488$
 & $3$&$488$ & $3$&$488$ \\
$1/\mu_1^{(\mathrm{LS})}$ &(fm)&
 $0$&$25$ & $0$&$25$ & $0$&$25$ & $0$&$25$ & $0$&$25$ \\
$t_1^{(\mathrm{LSE})}$ &(MeV)& $-5101$& & $-9181$&$8$
 & $-10712$&$1$ & $-8671$&$7$ & $-11222$&$2$ \\
$t_1^{(\mathrm{LSO})}$ &(MeV)& $-1897$& & $-3414$&$6$
 & $-3983$&$7$ & $-3224$&$9$ & $-4173$&$4$  \\
$1/\mu_2^{(\mathrm{LS})}$ &(fm)&
 $0$&$40$ & $0$&$40$ & $0$&$40$ & $0$&$40$ & $0$&$40$ \\
$t_2^{(\mathrm{LSE})}$ &(MeV)& $-337$& & $-606$&$6$
 & $-707$&$7$ & $-572$&$9$ & $-741$&$4$ \\
$t_2^{(\mathrm{LSO})}$ &(MeV)& $-632$& & $-1137$&$6$
 & $-1327$&$2$ & $-1074$&$4$ & $-1390$&$4$ \\
$1/\mu_1^{(\mathrm{TN})}$ &(fm)&
 $0$&$40$ & $0$&$40$ & $0$&$40$ & $0$&$40$ & $0$&$40$ \\
$t_1^{(\mathrm{TNE})}$ &(MeV$\cdot$fm$^{-2}$)& $-1096$& & $-131$&$52$
 & $-1096$& & $0$& & $-1096$& \\
$t_1^{(\mathrm{TNO})}$ &(MeV$\cdot$fm$^{-2}$)& $244$& & $29$&$28$
 & $244$& & $0$& & $244$& \\
$1/\mu_2^{(\mathrm{TN})}$ &(fm)&
 $0$&$70$ & $0$&$70$ & $0$&$70$ & $0$&$70$ & $0$&$70$ \\
$t_2^{(\mathrm{TNE})}$ &(MeV$\cdot$fm$^{-2}$)& $-30$&$9$ & $-3$&$708$
 & $-30$&$9$ & $0$& & $-30$&$9$ \\
$t_2^{(\mathrm{TNO})}$ &(MeV$\cdot$fm$^{-2}$)& $15$&$6$ & $1$&$872$
 & $15$&$6$ & $0$& & $15$&$6$ \\
$\alpha^{(\mathrm{SE})}$ && \multicolumn{2}{c}{---} &
 \multicolumn{2}{c}{$1/3$} & \multicolumn{2}{c}{$1$} &
 \multicolumn{2}{c}{$1$} & \multicolumn{2}{c}{$1$} \\
$t_\rho^{(\mathrm{SE})}$ &(MeV$\cdot$fm$^3$)& $0$& & $181$&~~$^{a)}$
 & $220$& & $248$& & $126$& \\
$\alpha^{(\mathrm{TE})}$ && \multicolumn{2}{c}{---} &
 \multicolumn{2}{c}{$1/3$} & \multicolumn{2}{c}{$1/3$} &
 \multicolumn{2}{c}{$1/3$} & \multicolumn{2}{c}{$1/3$} \\
$t_\rho^{(\mathrm{TE})}$ &(MeV$\cdot$fm)& $0$& & $1139$&
 & $1198$& & $1142$& & $1147$& \\
\hline\hline
\end{tabular}\\
$^{a)}$ (MeV$\cdot$fm)
\end{center}
\end{table}

\section{Properties of nuclear matter
at and around saturation point\label{sec:NMprop}}

We first view properties of the infinite nuclear matter
that are predicted by the semi-realistic $NN$ interactions.
In the HF approximation,
energy of the nuclear matter can be expressed
by the following variables:
\begin{eqnarray}
 \rho &=& {\displaystyle\sum_{\sigma\tau}} \rho_{\tau\sigma}\,,
  \nonumber\\
 \eta_s &=& \frac{{\displaystyle\sum_{\sigma\tau}}
  \sigma\rho_{\tau\sigma}}{\rho}
  ~=~ \frac{\rho_{p\uparrow}-\rho_{p\downarrow}+\rho_{n\uparrow}
  -\rho_{n\downarrow}}{\rho}\,, \nonumber\\
 \eta_t &=& \frac{{\displaystyle\sum_{\sigma\tau}}
  \tau\rho_{\tau\sigma}}{\rho}
  ~=~ \frac{\rho_{p\uparrow}+\rho_{p\downarrow}-\rho_{n\uparrow}
  -\rho_{n\downarrow}}{\rho}\,, \nonumber\\
 \eta_{st} &=& \frac{{\displaystyle\sum_{\sigma\tau}}
  \sigma\tau\rho_{\tau\sigma}}{\rho}
  ~=~ \frac{\rho_{p\uparrow}-\rho_{p\downarrow}-\rho_{n\uparrow}
  +\rho_{n\downarrow}}{\rho}\,.
\end{eqnarray}
$\rho_{\tau\sigma}$ ($\tau=p,n$ and $\sigma=\uparrow,\downarrow$,
which are sometimes substituted by $\pm 1$ without confusion)
stands for densities depending on the spin and the isospin,
and is related to the Fermi momentum $k_{\mathrm{F}\tau\sigma}$ by
\begin{equation}
 \rho_{\tau\sigma} = \frac{1}{6\pi^2} k_{\mathrm{F}\tau\sigma}^3\,.
\end{equation}
In the interaction of Eq.~(\ref{eq:effint}),
only $v^{(\mathrm{C})}+v^{(\mathrm{DD})}$
contributes to the energy of the uniform nuclear matter.
Formulas to calculate the nuclear matter energy and its derivatives
for given $k_{\mathrm{F}\tau\sigma}$ have been derived
in Ref.~\cite{ref:Nak03}.
Note that, even when superfluidity makes the nuclear matter energy
somewhat lower,
it is not much different from the energy in the HF approximation.

The spin-saturated symmetric matter is
characterized by $\eta_s=\eta_t=\eta_{st}=0$,
for which we denote $k_{\mathrm{F}\tau\sigma}$
simply by $k_\mathrm{F}$.
The minimum of the energy per nucleon $\mathcal{E}=E/A$,
given by
\begin{equation}
 \left.\frac{\partial\mathcal{E}}{\partial\rho}\right\vert_0
 =\left.\frac{\partial\mathcal{E}}{\partial k_\mathrm{F}}\right\vert_0
 =0\,,
\end{equation}
defines the saturation density $\rho_0$
(equivalently, $k_{\mathrm{F}0}$) and energy $\mathcal{E}_0$.
The expression $~\vert_0$ indicates evaluation at the saturation point.
As well as $\rho_0$ and $\mathcal{E}_0$,
second derivatives of $\mathcal{E}$ carry
basic information of the effective $NN$ interaction.
Two of the curvatures of $\mathcal{E}$ at the saturation point
are called incompressibility and volume symmetry energy,
\begin{equation}
 \mathcal{K} = k_\mathrm{F}^2 \left.\frac{\partial^2\mathcal{E}}
  {\partial k_\mathrm{F}^2}\right\vert_0
 = 9\rho^2 \left.\frac{\partial^2\mathcal{E}}{\partial\rho^2}
       \right\vert_0\,,\quad
 a_t = \left. \frac{1}{2} \frac{\partial^2\mathcal{E}}{\partial\eta_t^2}
	\right\vert_0\,,
\end{equation}
and are related to the Landau-Migdal (LM) parameters $f_0$ and $f'_0$ as
\begin{equation}
 \mathcal{K} = \frac{3k_{\mathrm{F}0}^2}{M^\ast_0}(1+f_0)\,,\quad
 a_t = \frac{k_{\mathrm{F}0}^2}{6M^\ast_0}(1+f'_0)\,,
\end{equation}
where $M^\ast_0$ represents the effective mass ($k$-mass)
at the saturation point.
See Ref.~\cite{ref:Nak03} for definition of the LM parameters.
The other curvatures of $\mathcal{E}$
with respect to $\eta_s$ and $\eta_{st}$,
denoted by $a_s$ and $a_{st}$, are defined analogously to $a_t$
and are expressed in terms of the LM parameters
$g_0$ and $g'_0$~\cite{ref:Nak03}.
The $k$-mass is defined by a derivative of the s.p. energy
$\varepsilon(\mathbf{k}\sigma\tau)$:
\begin{equation}
 \left.\frac{\partial\varepsilon(\mathbf{k}\sigma\tau)}{\partial k}
 \right\vert_0 = \frac{k_{\mathrm{F}0}}{M^\ast_0}\,,
\label{eq:M*}\end{equation}
and is connected to the LM parameter $f_1$ by
\begin{equation}
 \frac{M^\ast_0}{M} = 1+\frac{1}{3}f_1\,.
\end{equation}
In addition, density-dependence of the symmetry energy,
which is represented by a third derivative of $\mathcal{E}$ as
\begin{equation} \mathcal{L}_t = \left.\frac{1}{2}k_\mathrm{F}
 \frac{\partial^3\mathcal{E}}{\partial k_\mathrm{F}\,\partial\eta_t^2}
  \right\vert_0 = \left.\frac{3}{2}\rho
 \frac{\partial^3\mathcal{E}}{\partial\rho\,\partial\eta_t^2}
  \right\vert_0\,,
\end{equation}
is under interest in relevance to structure
of the neutron star crust~\cite{ref:OI07}.
These quantities calculated from the new semi-realistic interactions
are tabulated in Tables~\ref{tab:NMsat}.
We here set $M=(M_p+M_n)/2$,
where $M_p$ ($M_n$) is the measured mass of a proton
(a neutron)~\cite{ref:PDG06}.
For comparison,
the values obtained by the D1S parameter-set~\cite{ref:D1S}
of the Gogny interaction and from M3Y-P2 are also displayed.

\begin{table}
\begin{center}
\caption{Nuclear matter properties at the saturation point.
\label{tab:NMsat}}
\begin{tabular}{ccrrrrr}
\hline\hline
&&~~~~D1S~~ &~~M3Y-P2 &~~M3Y-P3 &~~M3Y-P4 &~~M3Y-P5 \\ \hline
$k_{\mathrm{F}0}$ & (fm) & $1.342$~~& $1.340$~~
 & $1.340$~~& $1.340$~~& $1.340$~~\\
$\mathcal{E}_0$ & (MeV) & $-16.01$~~& $-16.14$~~
 & $-16.51$~~& $-16.13$~~& $-16.12$~~\\
$\mathcal{K}$ & (MeV) & $202.9$~~& $220.4$~~
 & $245.8$~~& $235.3$~~& $235.6$~~\\
$M^\ast_0/M$ && $0.697$~~& $0.652$~~
 & $0.658$~~& $0.665$~~& $0.629$~~\\
$a_t$ & (MeV) & $31.12$~~& $30.61$~~
 & $29.75$~~& $28.71$~~& $29.59$~~\\
$a_s$ & (MeV) & $26.18$~~& $21.19$~~
 & $20.17$~~& $15.61$~~& $19.56$~~\\
$a_{st}$ & (MeV) & $29.13$~~& $38.19$~~
 & $36.45$~~& $39.89$~~& $41.01$~~\\
$\mathcal{L}_t$ & (MeV) & $22.44$~~& $27.98$~~
 & $25.30$~~& $17.87$~~& $24.63$~~\\
\hline\hline
\end{tabular}
\end{center}
\end{table}

Related to the global systematics of the binding energies and the radii,
$k_\mathrm{F0}\approx 1.33-1.34\,\mathrm{fm}^{-1}$
and $\mathcal{E}_0\approx -16\,\mathrm{MeV}$ have been established
empirically.
Although the M3Y-P3 interaction yields deeper $\mathcal{E}_0$
than the other interactions,
it is still within the range of ambiguity in extracting the volume energy
from the experimental data~\cite{ref:Kir08}.
In practice, M3Y-P3 does not yield overbinding
for any of the doubly magic nuclei presented in Sec.~\ref{sec:DMprop}.

For the incompressibility,
$\mathcal{K}\approx 240\,\mathrm{MeV}$ is extracted
from the experimental data~\cite{ref:SKC06}.
The $k$-mass is empirically known
to be $M_0^\ast\approx (0.6-0.7)M$~\cite{ref:Mahaux}.
The volume symmetry energy $a_t$ is important
in reproducing global trend of the binding energies
for the $Z\ne N$ nuclei,
and from empirical viewpoints $a_t\approx 30\,\mathrm{MeV}$
seems appropriate~\cite{ref:Dan03}.
These are fulfilled reasonably well
in all the new parameter-sets M3Y-P3 to P5.
The choice $\alpha^{(\mathrm{SE})}=1$
contributes to the slightly higher $\mathcal{K}$
in M3Y-P3 to P5 than in D1S and M3Y-P2.

Global characters of the spin and isospin responses
are customarily discussed in terms of the LM parameters.
By using the formulas given in Ref.~\cite{ref:Nak03},
we evaluate the LM parameters for the new semi-realistic interactions,
as shown in Table~\ref{tab:LM}.
It has been known that $g_0$ is small
while $g'_0$ is relatively large ($\approx 1$)~\cite{ref:g'0}.
Although the LM parameters should eventually be checked
by corresponding excitation modes in actual nuclei,
which is beyond the scope of this paper,
all the semi-realistic M3Y-type interactions
seem to have reasonable characters
on the spin and isospin channels.
Not necessarily true for phenomenological interactions such as D1S,
this may be linked to the microscopic origin of the interactions.
In particular, $v^{(\mathrm{C})}_\mathrm{OPEP}$
carries about half of $g'_0$ in the results
of the M3Y-type interactions~\cite{ref:Nak03}.

\begin{table}
\begin{center}
\caption{Landau-Migdal parameters at the saturation point.
\label{tab:LM}}
\begin{tabular}{cr@{.}lr@{.}lr@{.}lr@{.}lr@{.}l}
\hline\hline
\hspace*{1cm} & \multicolumn{2}{c}{~~D1S~~~~} &
 \multicolumn{2}{c}{M3Y-P2~~} & \multicolumn{2}{c}{M3Y-P3~~} &
 \multicolumn{2}{c}{M3Y-P4~~} & \multicolumn{2}{c}{M3Y-P5~~} \\ \hline
$f_0$ & $-0$&$369$ & $-0$&$357$ & $-0$&$276$ & $-0$&$300$ & $-0$&$336$ \\
$f_1$ & $-0$&$909$ & $-1$&$044$ & $-1$&$027$ & $-1$&$005$ & $-1$&$112$ \\
$f_2$ & $-0$&$558$ & $-0$&$436$ & $-0$&$355$ & $-0$&$429$ & $-0$&$367$ \\
$f_3$ & $-0$&$157$ & $-0$&$210$ & $-0$&$184$ & $-0$&$210$ & $-0$&$182$ \\
 \hline
$f'_0$ & $0$&$743$ & $0$&$607$ & $0$&$578$ & $0$&$538$ & $0$&$502$ \\
$f'_1$ & $0$&$470$ & $0$&$635$ & $0$&$670$ & $0$&$797$ & $0$&$692$ \\
$f'_2$ & $0$&$342$ & $0$&$245$ & $0$&$271$ & $0$&$286$ & $0$&$267$ \\
$f'_3$ & $0$&$100$ & $0$&$096$ & $0$&$104$ & $0$&$106$ & $0$&$100$ \\
 \hline
$g_0$ & $0$&$466$ & $0$&$113$ & $0$&$070$ & $-0$&$164$ & $-0$&$007$ \\
$g_1$ & $-0$&$184$ & $0$&$273$ & $0$&$214$ & $0$&$374$ & $0$&$299$ \\
$g_2$ & $0$&$245$ & $0$&$162$ & $0$&$160$ & $0$&$190$ & $0$&$178$ \\
$g_3$ & $0$&$091$ & $0$&$078$ & $0$&$079$ & $0$&$085$ & $0$&$081$ \\
 \hline
$g'_0$ & $0$&$631$ & $1$&$006$ & $0$&$933$ & $1$&$136$ & $1$&$081$ \\
$g'_1$ & $0$&$610$ & $0$&$202$ & $0$&$213$ & $0$&$109$ & $0$&$087$ \\
$g'_2$ & $-0$&$038$ & $0$&$040$ & $0$&$063$ & $0$&$016$ & $0$&$029$ \\
$g'_3$ & $-0$&$036$ & $-0$&$002$ & $0$&$005$ & $-0$&$008$ & $-0$&$002$ \\
\hline\hline
\end{tabular}
\end{center}
\end{table}

\begin{figure}
\includegraphics[scale=0.9]{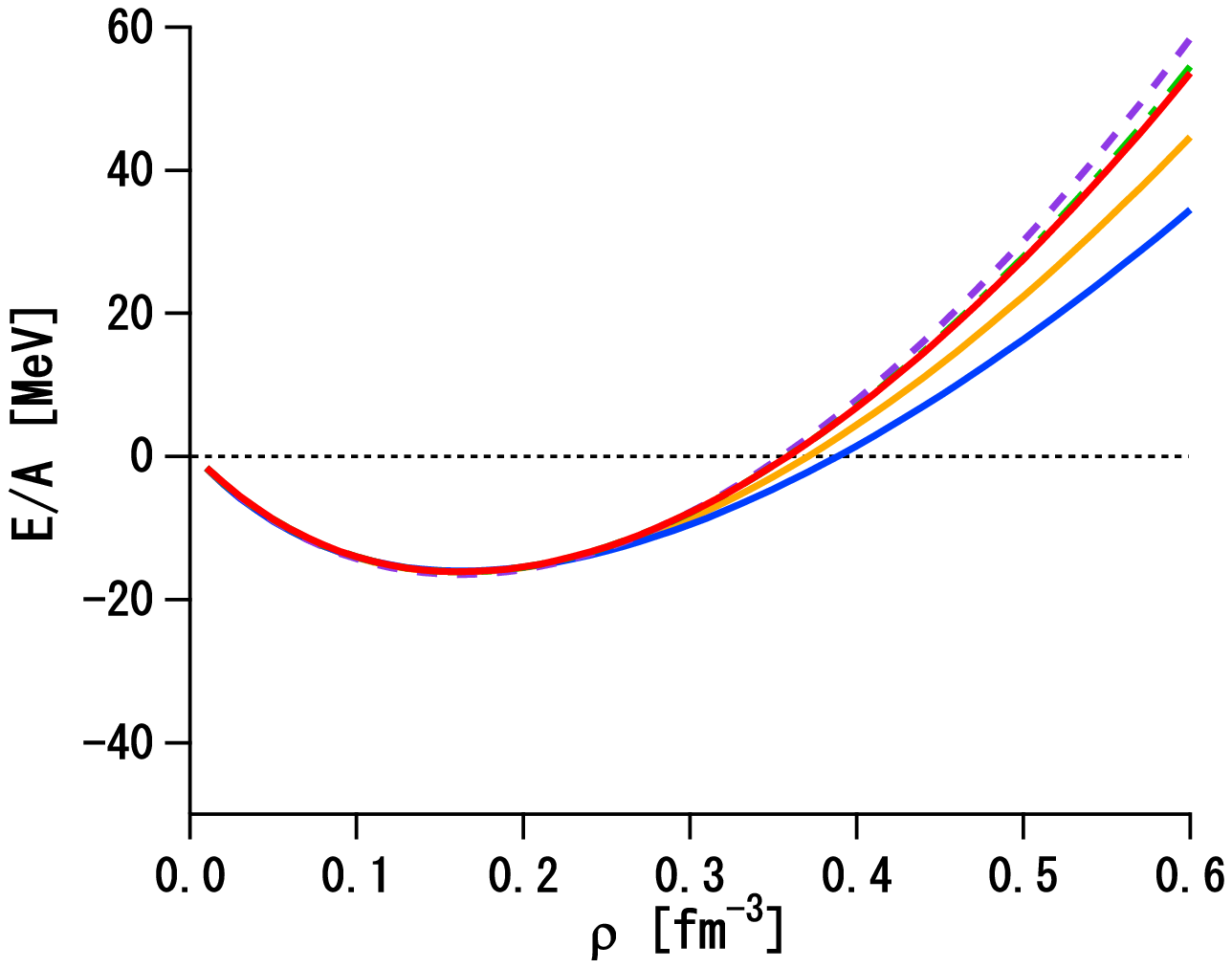}
\caption{Energy per nucleon $\mathcal{E}=E/A$
in the symmetric nuclear matter for several effective interactions.
Purple dashed, green dot-dashed and red solid lines
represent the results with the M3Y-P3, P4 and P5
interactions, respectively.
Those with the M3Y-P2 and D1S interactions
are also displayed for comparison
by orange and blue solid lines.
\label{fig:NME_M3Ya}}
\end{figure}

Figure~\ref{fig:NME_M3Ya} illustrates $\mathcal{E}(\rho)$
for the spin-saturated symmetric nuclear matter
obtained from the M3Y-type and the D1S interactions.
As pointed out in Ref.~\cite{ref:Nak03},
difference among the saturating forces
is not large at $\rho\lesssim \rho_0$.
At relatively high density ($\rho\gtrsim 0.3\,\mathrm{fm}^{-3}$),
the M3Y-P3 to P5 interactions have higher $\mathcal{E}$
than M3Y-P2 and D1S,
reflecting higher $\mathcal{K}$.
$\mathcal{E}(\rho)$ of M3Y-P4 and P5 is close to each other
even at $\rho\approx 0.6\,\mathrm{fm}^{-3}(\approx 4\rho_0)$.

\begin{figure}
\includegraphics[scale=0.6]{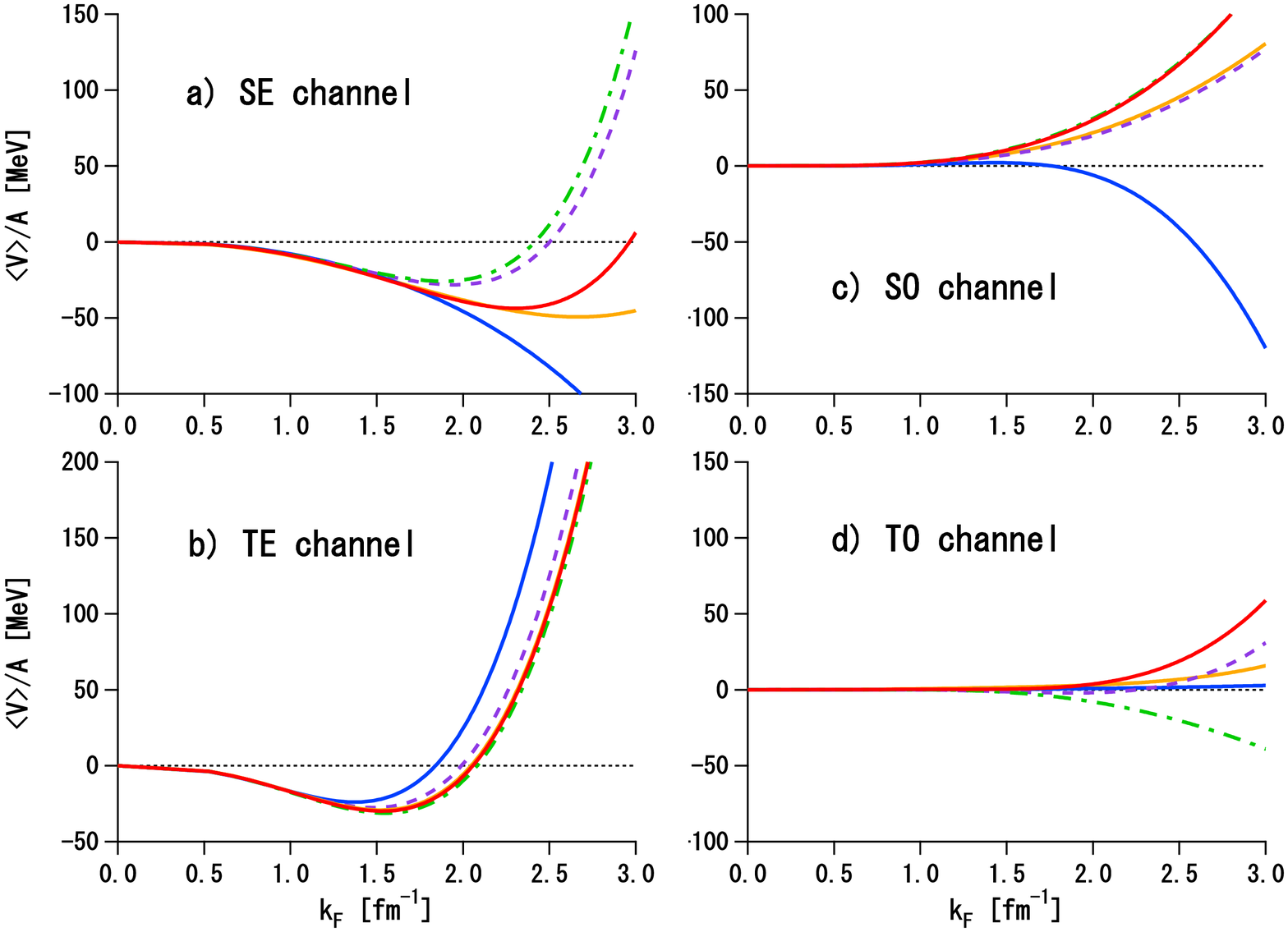}
\caption{Contribution of the SE, TE, SO and TO channels to $\mathcal{E}$.
See Fig.~\protect\ref{fig:NME_M3Ya} for conventions.
\label{fig:NMV_all}}
\end{figure}

In Fig.~\ref{fig:NMV_all},
contributions of the SE, TE, SO and TO channels
in $v^{(\mathrm{C})}+v^{(\mathrm{DD})}$
to $\mathcal{E}$ of the symmetric matter
are shown as a function of $k_\mathrm{F}$.
The contribution of the TE and the SO channels in M3Y-P4
is hard to be distinguished from that in M3Y-P5.
So is the contribution of the TE channel in M3Y-P2.
The TE channel takes a minimum
at $k_\mathrm{F}\approx 1.5\,\mathrm{fm}^{-1}$,
primarily responsible for the saturation
at $k_\mathrm{F}=k_{\mathrm{F}0}\approx 1.3\,\mathrm{fm}^{-1}$.
Both the SO and the TO channels do not contribute to $\mathcal{E}$
significantly at $\rho\lesssim\rho_0$
(\textit{i.e.} $k_\mathrm{F}\lesssim k_{\mathrm{F}0}$).
While the SO channel becomes attractive
in the D1S interaction,
it is repulsive in the M3Y-type interactions at $\rho>\rho_0$.
The TO channel is repulsive in M3Y-P3 and P5,
while attractive in M3Y-P4,
although the attraction in M3Y-P4 is not so strong
as to cause spin polarization in the pure neutron matter
up to $\rho=1.3\,\mathrm{fm}^{-3}(\approx 8\rho_0)$.
Remember that the odd channels in M3Y-P3 are unchanged
from M3Y-P0.

\begin{figure}
\includegraphics[scale=0.9]{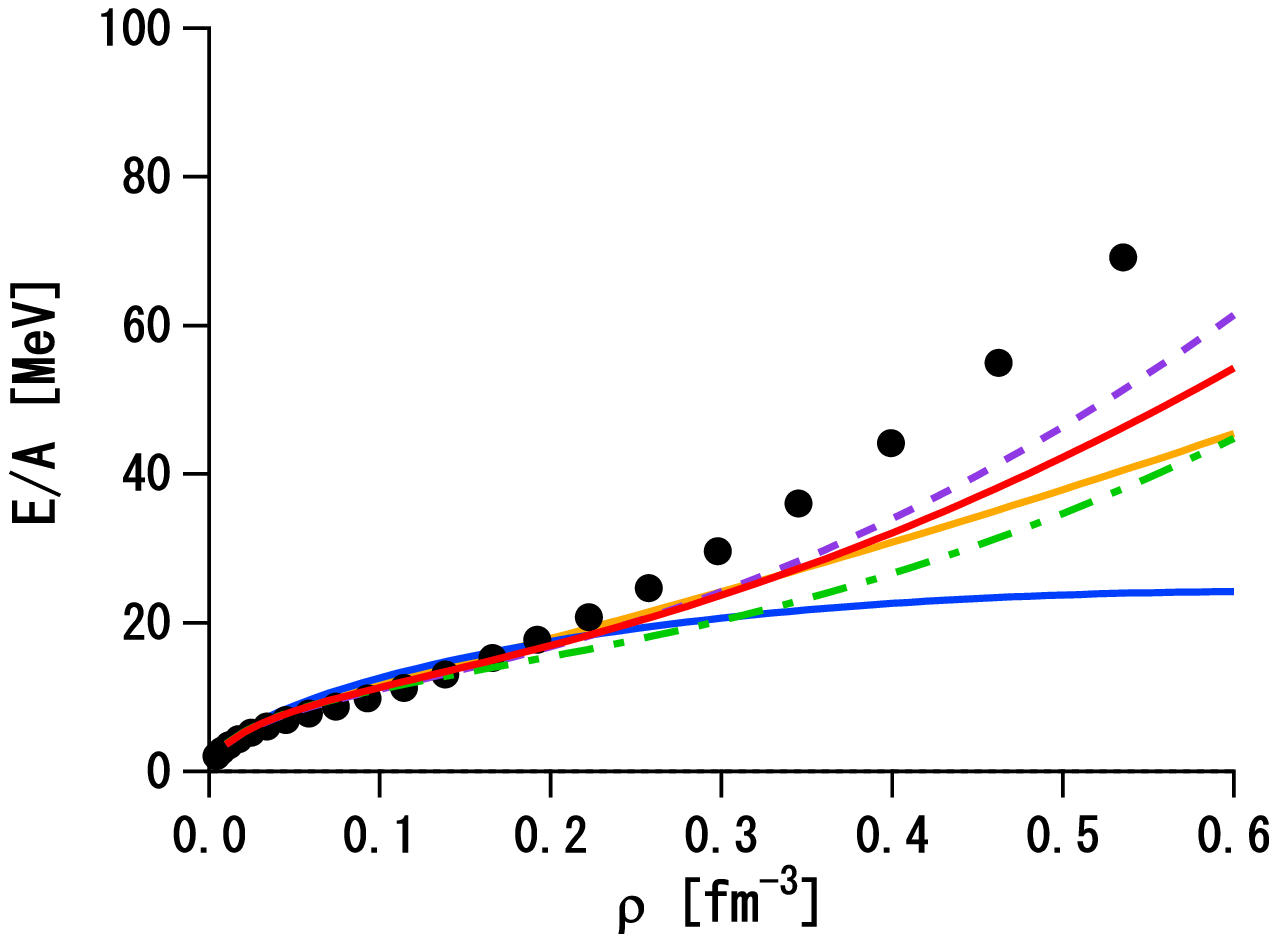}
\caption{Energy per nucleon $\mathcal{E}=E/A$
in the neutron matter for several effective interactions.
Circles are the results of Ref.~\protect\cite{ref:FP81}.
See Fig.~\protect\ref{fig:NME_M3Ya} for the other conventions.
\label{fig:NME_M3Yc}}
\end{figure}

Energy per nucleon in the spin-saturated neutron matter
(\textit{i.e.} $\eta_t=-1$) is presented in Fig.~\ref{fig:NME_M3Yc}.
The result from a microscopic calculation in Ref.~\cite{ref:FP81}
is also shown as a reference.
The unphysical behavior at high $\rho$ in the D1S result,
which comes from the absence of density-dependence in the SE channel,
was pointed out in Refs.~\cite{ref:NS02,ref:Nak03}.
The present M3Y-P3 to P5 interactions have
relatively strong $\rho$-dependence
at high $\rho$ for the neutron matter,
if compared to M3Y-P2 and D1S.
This originates in $\alpha^{(\mathrm{SE})}(=1)$,
and tends to make $\mathcal{E}(\rho)$
closer to the result of Ref.~\cite{ref:FP81}.

\section{Properties of doubly magic nuclei\label{sec:DMprop}}

We next turn to doubly magic nuclei,
for which the spherical HF approach is expected
to be a good approximation.

To all the following calculations of finite nuclei,
we apply the recently developed algorithm
based on the Gaussian expansion method
(GEM)~\cite{ref:NS02,ref:Nak06}.
In this method we employ the s.p. bases of
\begin{eqnarray} \varphi_{\nu\ell jm}(\mathbf{r})
= R_{\nu\ell j}(r)[Y^{(\ell)}(\hat{\mathbf{r}})\chi_\sigma]^{(j)}_m\,;
\quad
R_{\nu\ell j}(r) = \mathcal{N}_{\nu\ell j}\,
r^\ell \exp(-\nu r^2)\,,
\label{eq:basis} \end{eqnarray}
apart from the isospin index.
Here $Y^{(\ell)}(\hat{\mathbf{r}})$ expresses the spherical harmonics
and $\chi_\sigma$ the spin wave function.
The parameter $\nu=\nu_\mathrm{r}+i\nu_\mathrm{i}$
indicates a complex number
corresponding to the range of the Gaussian.
Irrespective to nuclide,
we adopt the following basis parameters~\cite{ref:Nak08}:
\begin{equation}
\nu_\mathrm{r}=\nu_0\,b^{-2n}\,,\quad
\left\{\begin{array}{ll}\nu_\mathrm{i}=0 & (n=0,1,\cdots,5)\\
{\displaystyle\frac{\nu_\mathrm{i}}{\nu_\mathrm{r}}
=\pm\frac{\pi}{2}} & (n=0,1,2)\end{array}\right.\,,
 \label{eq:basis-param}
\end{equation}
with $\nu_0=(2.40\,\mathrm{fm})^{-2}$ and $b=1.25$
for each $(\ell,j)$.
It is notable that,
without parameters specific to mass number or nuclide,
a single set of the GEM bases is applicable
to wide range of the nuclear mass table~\cite{ref:Nak08}.
The Hamiltonian is $H=H_N+V_C-H_\mathrm{c.m.}$,
where $V_C$ and $H_\mathrm{c.m.}$ represent the Coulomb interaction
and the center-of-mass (c.m.) Hamiltonian,
while $H_N$ has been given in Eq.~(\ref{eq:H_N}).
The exchange term of $V_C$ is treated exactly,
in the same manner as the nuclear force $V_N$.
Both the one- and the two-body terms of $H_\mathrm{c.m.}$
are subtracted before iteration.

The calculated binding energies and rms matter radii
of several doubly magic nuclei
are displayed in Table~\ref{tab:DMprop}.
The results of the new semi-realistic interactions are
compared with those of D1S and M3Y-P2
as well as with the experimental data.
Influence of the c.m. motion on the matter radii
is subtracted in a similar manner
to the c.m. energies~\cite{ref:Nak03}.
The binding energies of these nuclei obtained from D1S (M3Y-P2)
are in agreement with the measured values
within the $3\,\mathrm{MeV}$ ($5\,\mathrm{MeV}$) accuracy.
Though the accuracy is slightly worse,
the new interactions also reproduce the binding energies
moderately well.
M3Y-P3 yields underbinding by about $9-17\,\mathrm{MeV}$
except for $^{208}$Pb.
For M3Y-P4 and P5,
maximum deviation in the binding energies shown in Table~\ref{tab:DMprop}
is $\sim 7\,\mathrm{MeV}$.
Since correlations due to the residual interaction could influence,
we do not take this deviation seriously at the present stage.
The rms matter radii of these nuclei calculated
from the semi-realistic interactions
are comparable to those from the D1S interaction,
in fair agreement with the data.

\begin{table}
\begin{center}
\caption{Binding energies and rms matter radii
 of several doubly magic nuclei.
 Experimental data are taken
 from Refs.~\protect\cite{ref:mass,ref:O16-rad,ref:O24-rad,ref:rad}.
\label{tab:DMprop}}
\begin{tabular}{cccrrrrrr}
\hline\hline
&&&~~~Exp.~~&~~~~~D1S~~&~~M3Y-P2~&~~M3Y-P3~&~~M3Y-P4~&~~M3Y-P5~\\ \hline
$^{16}$O & $-E$ &(MeV)&
 $127.6$ & $129.5$ & $127.2$ & $118.6$ & $126.3$ & $126.1$ \\
& $\sqrt{\langle r^2\rangle}$ &(fm)&
 $2.61$ & $2.61$ & $2.61$ & $2.65$ & $2.60$ & $2.59$ \\
$^{24}$O & $-E$ &(MeV)&
 $168.5$ & $168.6$ & $165.7$ & $158.2$ & $164.0$ & $166.7$ \\
& $\sqrt{\langle r^2\rangle}$ &(fm)&
 $3.19$ & $3.01$ & $3.06$ & $3.08$ & $3.04$ & $3.03$ \\
$^{40}$Ca & $-E$ &(MeV)&
 $342.1$ & $344.6$ & $338.8$ & $325.2$ & $337.0$ & $335.1$ \\
& $\sqrt{\langle r^2\rangle}$ &(fm)&
 $3.47$ & $3.37$ & $3.38$ & $3.42$ & $3.37$ & $3.37$ \\
$^{48}$Ca & $-E$ &(MeV)&
 $416.0$ & $416.8$ & $411.9$ & $401.0$ & $409.4$ & $414.1$ \\
& $\sqrt{\langle r^2\rangle}$ &(fm)&
 $3.57$ & $3.51$ & $3.53$ & $3.56$ & $3.52$ & $3.50$ \\
$^{90}$Zr & $-E$ &(MeV)&
 $783.9$ & $785.9$ & $779.4$ & $767.9$ & $775.1$ & $779.8$ \\
& $\sqrt{\langle r^2\rangle}$ &(fm)&
 $4.32$ & $4.24$ & $4.25$ & $4.27$ & $4.24$ & $4.23$ \\
$^{132}$Sn & $-E$ &(MeV)&
 $1102.9$ & $1104.1$ & $1099.0$ & $1089.3$ & $1095.7$ & $1098.4$ \\
& $\sqrt{\langle r^2\rangle}$ &(fm)&
 --- & $4.77$ & $4.79$ & $4.81$ & $4.77$ & $4.76$ \\
$^{208}$Pb & $-E$ &(MeV)&
 $1636.4$ & $1639.0$ & $1636.5$ & $1635.2$ & $1632.1$ & $1633.2$ \\
& $\sqrt{\langle r^2\rangle}$ &(fm)&
 $5.49$ & $5.51$ & $5.54$ & $5.55$ & $5.51$ & $5.51$ \\
\hline\hline
\end{tabular}
\end{center}
\end{table}

\begin{figure}
\includegraphics[scale=1.0]{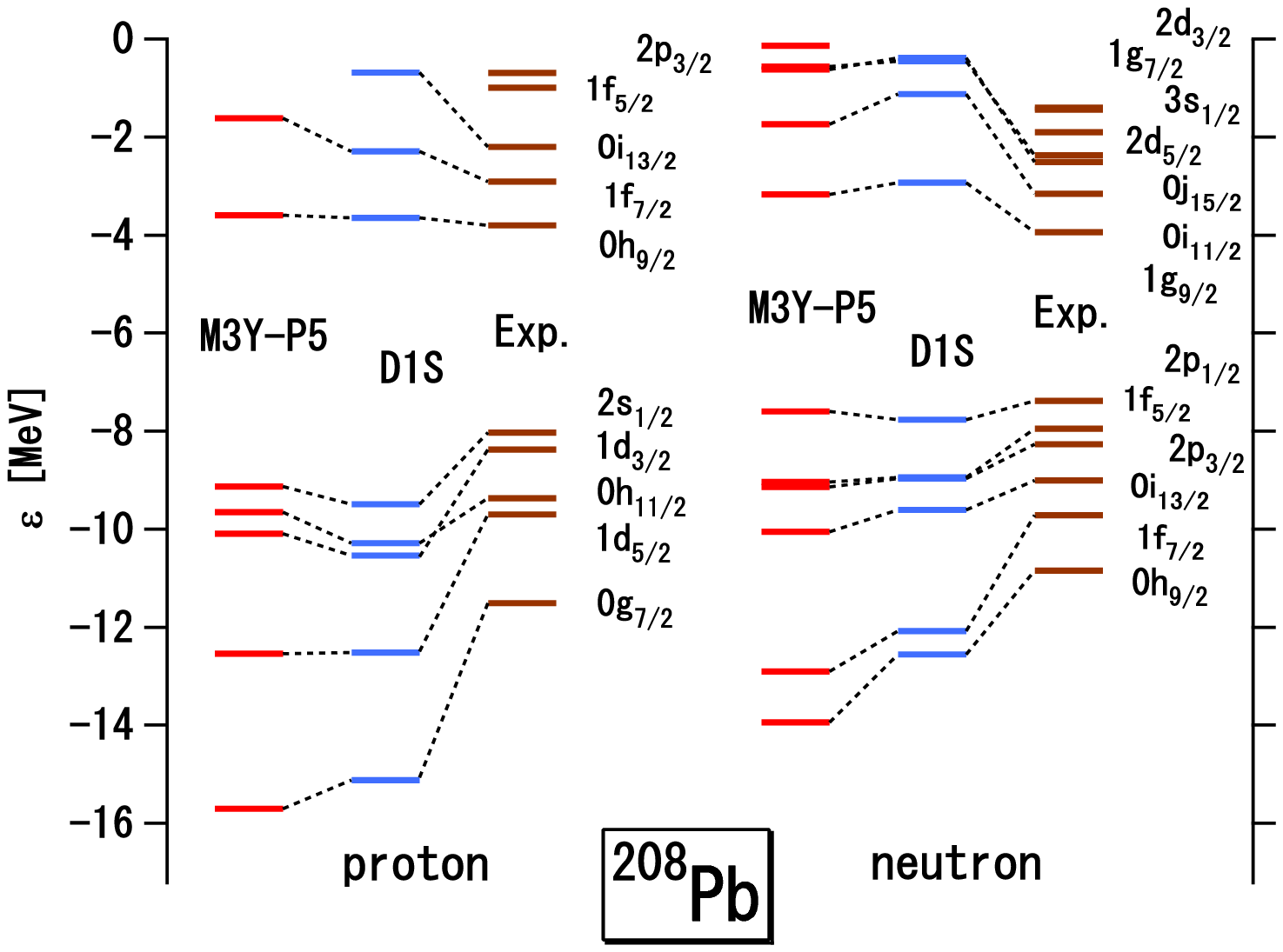}
\caption{Single-particle energies for $^{208}$Pb.
Experimental values are extracted
from Refs.~\protect\cite{ref:mass,ref:TI}.
\label{fig:Pb_spe}}
\end{figure}

The s.p. levels in $^{208}$Pb are depicted in Fig.~\ref{fig:Pb_spe}.
The levels obtained from M3Y-P5 are compared
with those from D1S and the experimental levels.
M3Y-P3 and P4 give s.p. levels similar to,
though slightly different from, those of M3Y-P5.
The experimental s.p. energies are obtained from the levels
of the neighboring nuclei; $^{207,209}$Pb, $^{207}$Tl and $^{209}$Bi.
In the HF results,
the overall level spacing is relevant to $M^\ast_0$
shown in Table~\ref{tab:NMsat}.
In the usual HF calculations we have larger level spacing
than in the data,
and it is not (should not be) remedied until correlations
due to the residual interaction (or the $\omega$-mass)
are taken into account~\cite{ref:Mahaux}.
We thus confirm that the present interactions yield
as plausible s.p. levels as D1S does.

\begin{figure}
\includegraphics[scale=1.0]{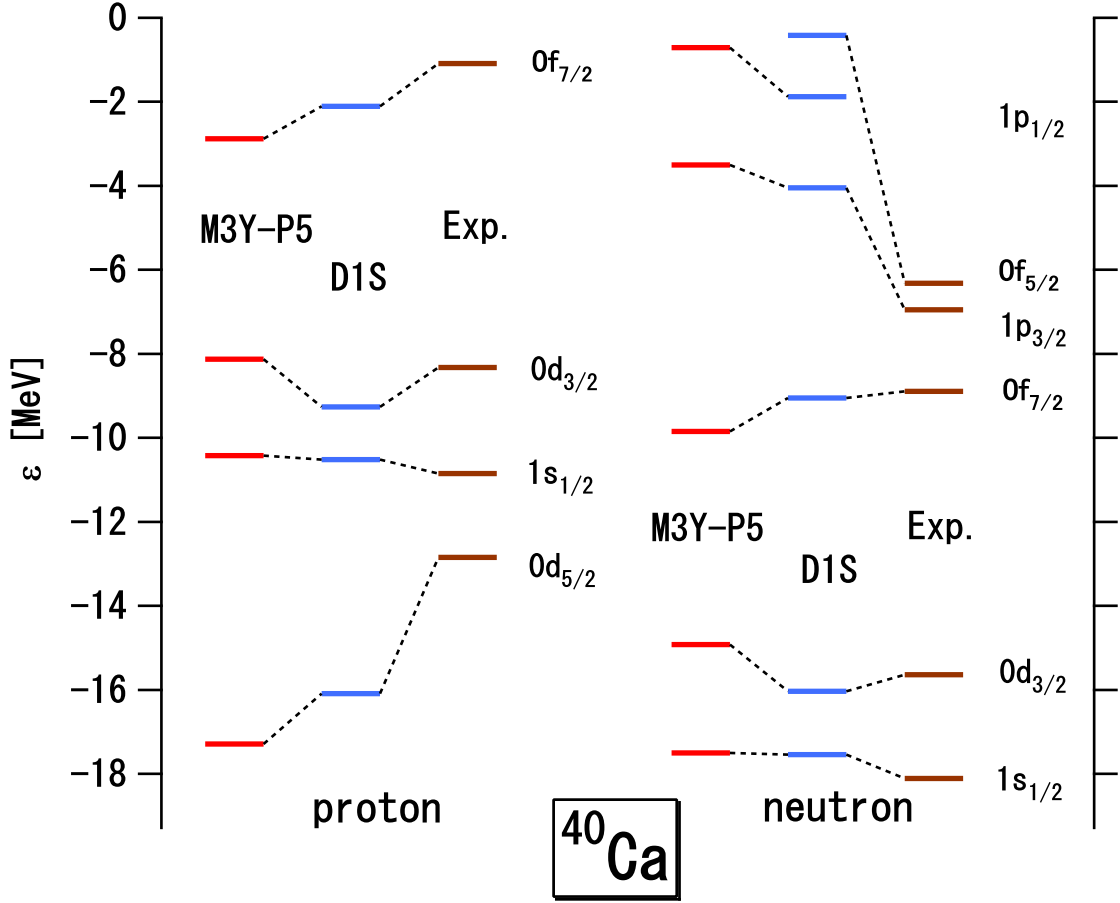}
\caption{Single-particle energies for $^{40}$Ca.
Experimental values are extracted
from Refs.~\protect\cite{ref:mass,ref:TI}.
\label{fig:Ca_spe}}
\end{figure}

In Ref.~\cite{ref:LBB07},
it has been shown that the $Z=N=20$ shell gaps are narrowed
by the tensor force.
It is also true in the M3Y-type interactions.
The s.p. energy difference
$\varepsilon_\tau(0f_{7/2})-\varepsilon_\tau(0d_{3/2})$ ($\tau=p,n$)
in $^{40}$Ca obtained by the D1S interaction
is in good agreement with the experimental values,
both for protons and neutrons, as viewed in Fig.~\ref{fig:Ca_spe}.
While M3Y-P2 and P4 give almost the same size of the shell gaps
as D1S does,
we have narrower gaps in the HF calculations with M3Y-P3 and P5.
However, the shell gaps do not collapse by $v^{(\mathrm{TN})}$ of M3Y,
in contrast to the zero-range tensor force of Ref.~\cite{ref:LBB07}.
We still have $5.2\,\mathrm{MeV}$ ($7.7\,\mathrm{MeV}$) gap
for the proton (neutron) orbits with M3Y-P5.
These gaps are close to those obtained
from the tensor-free Skyrme interaction `T22' in Ref.~\cite{ref:LBB07}.
It is also worth commenting that, for M3Y-P5,
the octupole correlations significantly influence
the ground state of $^{40}$Ca,
as will be discussed elsewhere.
This can make the shell gap look wider,
having possibility to account for the observed gap.

\section{Pairing properties\label{sec:pairing}}

The M3Y-P2 interaction seems to have reasonable characters
in the HF regime,
as exemplified in Table~\ref{tab:DMprop}.
However, M3Y-P2 has too strong pair correlations,
indicating too strong attraction in the SE channel at low densities
though almost invisible in Fig.~\ref{fig:NMV_all}.
This character is inherited from the original M3Y interaction.
We have developed the M3Y-P3 to P5 parameter-sets
by taking the pairing properties into consideration.
In this section we shall show characters of the new interactions
with respect to the pairing.
We restrict ourselves to the pairing among like-nucleons, as usual.

We implement the spherical HFB calculations for finite nuclei,
using the GEM bases of Eqs.~(\ref{eq:basis},\ref{eq:basis-param})
together with the $\ell\leq 7$ truncation.
The blocked HFB calculations are applied to the odd-mass nuclei,
by assuming that a quasiparticle (q.p.) occupies
a specified spherical orbital.
When several q.p. levels lie closely in energy,
we compare the total energies by filling each q.p. level
and adopt the lowest-energy solution.

\subsection{Even-odd mass difference in Sn isotopes
  \label{subsec:pair-Sn}}

\begin{figure}
\includegraphics[scale=1.1]{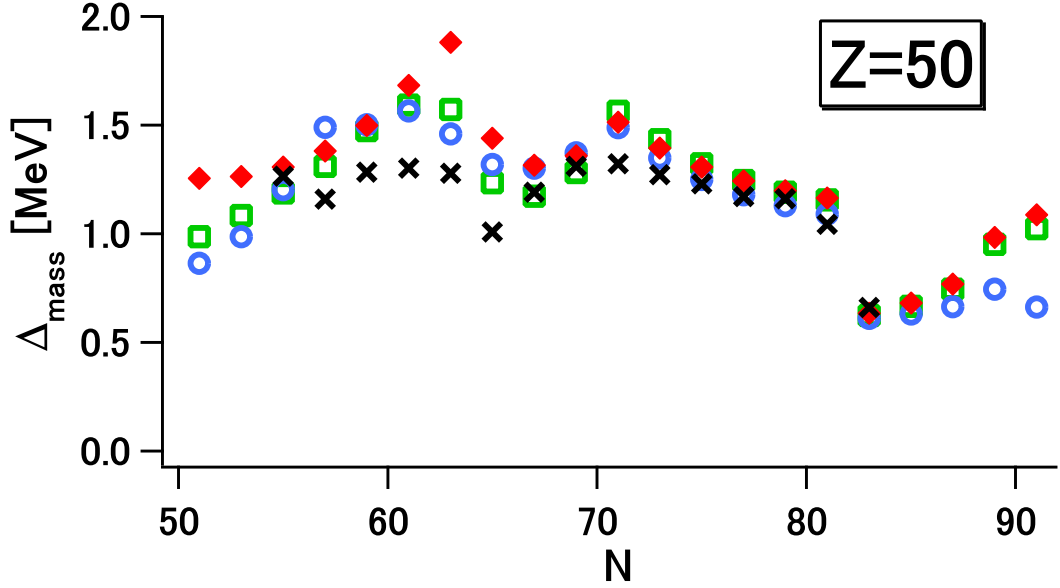}
\caption{Even-odd mass differences in the Sn isotopes,
$\Delta_\mathrm{mass}^{Z=50}(N)$.
The results of D1S, M3Y-P4 and M3Y-P5 are shown by blue open circles,
green open squares and red diamonds, respectively.
Experimental values, presented by black crosses,
are taken from Ref.~\protect\cite{ref:mass}.
\label{fig:Sn_Dmass}}
\end{figure}

The $t_n^{(\mathrm{SE})}$ ($n=1,2$)
and $t_\rho^{(\mathrm{SE})}$ parameters
of M3Y-P3 to P5 are adjusted to the even-odd mass differences
of the Sn isotopes with $66<N<80$.
For the mass difference we use the three-point formula
$\Delta_\mathrm{mass}^Z(N)=E(Z,N)-\frac{1}{2}\big[E(Z,N+1)+E(Z,N-1)\big]$,
with $Z=50$ and $N=\mathrm{odd}$.
The mass differences calculated with M3Y-P4 and P5
are displayed in Fig.~\ref{fig:Sn_Dmass},
in comparison with the experimental data and with those of D1S.
Though not shown to keep the figure viewable,
M3Y-P3 gives similar $\Delta_\mathrm{mass}^{Z=50}(N)$ to M3Y-P4.
The calculations are not fully convergent for the $\ell$ truncation.
Moreover,
the restoration of the particle-number conservation~\cite{ref:AER02}
and the non-spherical mean fields~\cite{ref:RBRM99} could influence
$\Delta_\mathrm{mass}^Z(N)$.
Each of them could vary $\Delta_\mathrm{mass}^Z(N)$
by up to a few hundred keV.
Not attempting fine tuning of the parameters,
we just point out that some of these effects
tend to compensate one another in the mass differences,
and that the new interactions give $\Delta_\mathrm{mass}^{Z=50}(N)$
to comparable accuracy to the D1S interaction in the same model space.

At $N\sim 50$, $64$ and $90$, we find that
the calculated mass differences depend on the interactions.
This is ascribed to interaction-dependence of the shell structure.
Irregularity and discrepancy at $N=63$ and $65$
should be relevant to the $N=64$ subshell.
In M3Y-P5 the subshell effect seems stronger
than in the other interactions.
At $N\sim 90$ all the M3Y-type interactions
yield larger mass differences than D1S.
This takes place because $n1f_{7/2}$ and $n2p_{3/2}$ well mix
due to the pairing, in the M3Y-type interactions.
At $N\sim 50$ M3Y-P5 yields larger mass difference
than the other interactions.
This is traced back to appreciable excitation
from $n1d_{5/2}$ to $n0g_{7/2}$,
which takes place since these two orbits are close in energy.
Possibly carrying information of the shell structure,
data on the masses in $N\sim 50$ and $N\sim 90$ will be of interest.

\subsection{Pairing gap in nuclear matter
  \label{subsec:pair-NM}}

We next view the pairing property in the nuclear matter
obtained from the new semi-realistic interactions.
In phenomenological studies using the Skyrme energy density functionals,
it has been argued~\cite{ref:DNR01}
whether and how much the pair correlations
are dominated by the nuclear surface region.
Results of the semi-realistic interactions for the nuclear matter
may provide certain information on this point.
However, we find that, in calculating the pairing properties,
the Yukawa function gives quite slow convergence
for the maximum momentum of the s.p. states.
In practice, even if we cut off the momentum at $k=50\,\mathrm{fm}^{-1}$,
which corresponds to $\varepsilon\approx 50\,\mathrm{GeV}$,
the pairing gap is not yet fully convergent.
It is impractical to include such high energy states
in calculations of finite nuclei.
As shown in the preceding subsection,
we have fixed the SE channel parameters
from the HFB calculations of the Sn nuclei
using the basis-parameters of Eq.~(\ref{eq:basis-param}).
High momentum components are automatically excluded in the basis set.
It will be natural to introduce a certain cut-off
in arguing the pairing in the nuclear matter,
and the cut-off should desirably be consistent with the basis set
of Eq.~(\ref{eq:basis-param}).

The basis set is composed of radial Gaussians,
whose Fourier transforms are again Gaussians in the momentum space.
We here consider a cut-off factor for the s.p. momentum space of
\begin{equation}
 g(k) = \theta(k_c-k) + \theta(k-k_c)\,
  \exp\!\left[-\big(\frac{k-k_c}{k_d}\big)^2\right]\,.
 \label{eq:cutoff}
\end{equation}
The measure in the $k$ integration is multiplied by $g(k)$.
Among the bases of Eq.~(\ref{eq:basis-param}),
the highest $k$ component is given by
the $\nu=\nu_0(1\pm\frac{\pi}{2}i)$ basis.
Since the Fourier transform of this basis is proportional to
$\exp\!\big[-k^2(1\pm\frac{\pi}{2}i)/4\nu_0(1+\frac{\pi^2}{4})\big]$,
$k\lesssim k_0=2\sqrt{\nu_0(1+\frac{\pi^2}{4})}
(\approx 1.55\,\mathrm{fm}^{-1})$ components
are well included in the set.
To be consistent with the basis set,
it will be reasonable to take $k_c\sim k_0$
for the nuclear matter calculation.
We here consider three cases, $(k_c,k_d)=(k_0,k_0)$, $(2k_0,k_0)$
and $(4\,\mathrm{fm}^{-1},0)$.
The last choice of the $k_d\rightarrow 0$ limit
indicates a sharp cut-off,
and $k_c=4\,\mathrm{fm}^{-1}$ approximately corresponds
to the maximum q.p. energy in the HFB calculations of the Sn nuclei.

\begin{figure}
\includegraphics[scale=1.1]{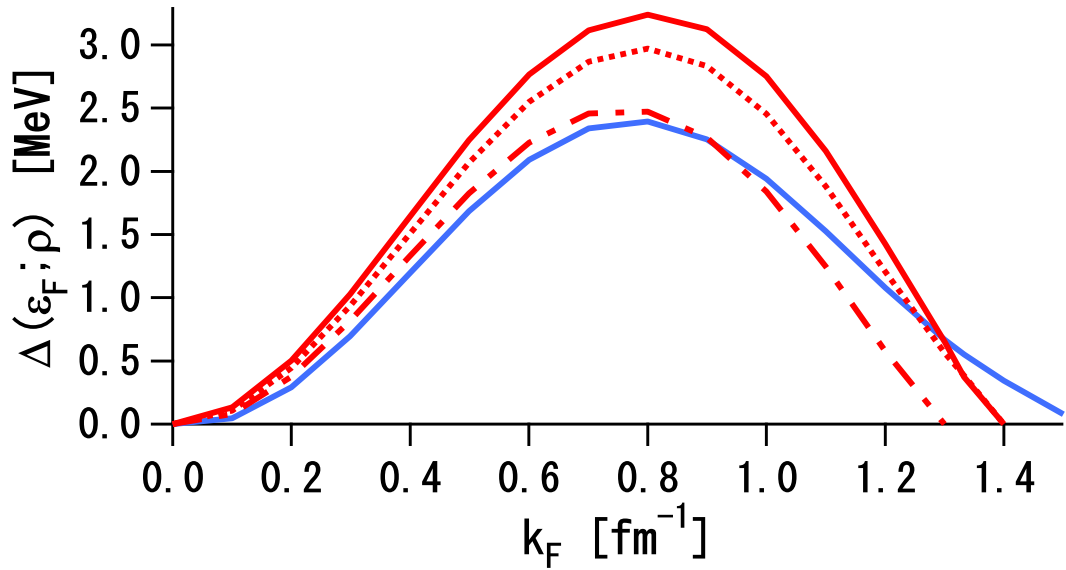}
\caption{Pairing gap at the Fermi level $\varepsilon_\mathrm{F}$
in the symmetric nuclear matter.
Red dot-dashed, solid and dotted curves are obtained from M3Y-P5
with the momentum cut-off $(k_c,k_d)=(k_0,k_0)$,
$(2k_0,k_0)$ and $(4\,\mathrm{fm}^{-1},0)$
in Eq.~(\protect\ref{eq:cutoff}), respectively.
Blue solid curve displays gap of the D1S interaction.
\label{fig:NMgap}}
\end{figure}

In Fig.~\ref{fig:NMgap},
the pairing gap at the Fermi energy,
which is obtained from the Bardeen-Cooper-Schrieffer (BCS) calculation
in the symmetric nuclear matter
using the method of Ref.~\cite{ref:NM-BCS},
is plotted as a function of $k_\mathrm{F}=(3\pi^2\rho/2)^{1/3}$.
The pairing among like-nucleons arises from the SE channel
in the effective interaction.
The result of the M3Y-P5 interaction is compared
with that of the Gogny D1S interaction.
The cut-off is not needed for the Gogny interaction,
and the present cut-off does not influence the D1S gap.
The pairing gaps of M3Y-P3 and P4 are similar to that of M3Y-P5.

The new semi-realistic interactions are not drastically different from D1S,
in respect to the pairing properties in the nuclear matter.
The gap has a peak at $k_\mathrm{F}\approx 0.8\,\mathrm{fm}^{-1}
\approx 0.6k_{\mathrm{F}0}$, namely at $\rho\approx 0.2\rho_0$,
for all cases.
However, the peak height and the behavior at $\rho>0.2\rho_0$
are different between the M3Y-type interactions and D1S.
With rapid decrease at $\rho>0.2\rho_0$,
the M3Y-type interactions have more surface-dominant pairing than D1S.
In D1S, the pair correlation may have sizable contribution
from the bulk,
even though it is stronger at the nuclear surface.
Note that the cut-off parameter does not influence
the nuclear matter pairing qualitatively,
as long as it is more or less harmonious with the basis-set
adopted in the calculations of finite nuclei.

\section{Neutron drip line\label{sec:n-drip}}

Prediction of the neutron drip line depends on effective interactions
to a certain degree.
In this section we compare location of the neutron drip line
predicted by the spherical HFB calculations
with the present semi-realistic interactions
and with the Gogny D1S interaction for the O, Ca and Ni isotopes.
Although complete description of the drip line
may require fine tuning of the parameters
as well as taking account of correlation effects,
it will be interesting to see
what is relevant to location of the drip line.

\subsection{$Z=8$ nuclei\label{subsec:drip-Z8}}

\begin{figure}
\includegraphics[scale=1.0]{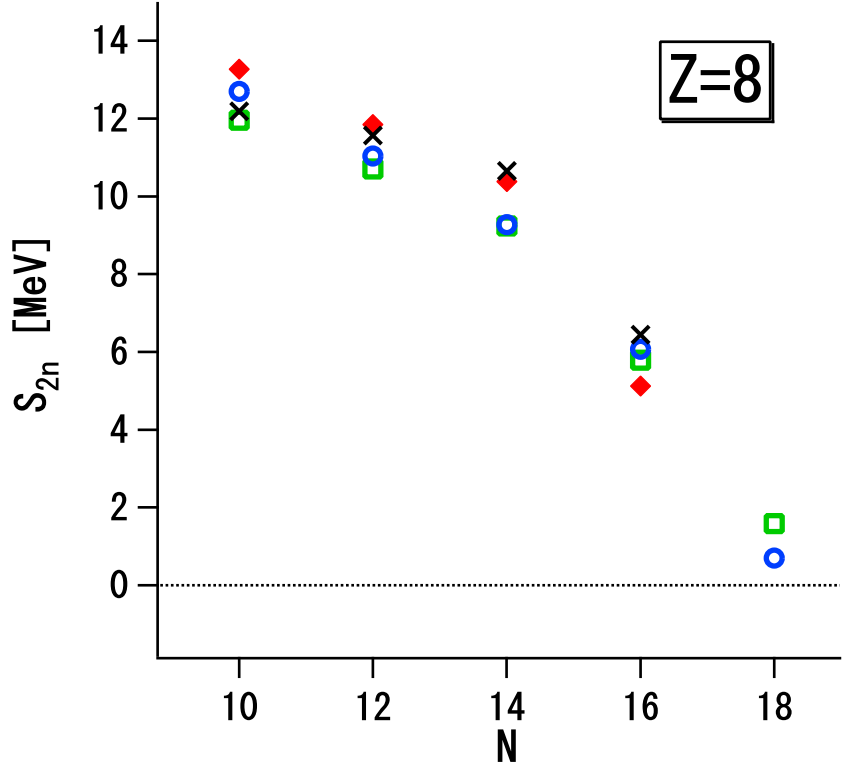}
\caption{$S_{2n}$ of the O isotopes ($N=\mbox{even}$).
The calculations are performed in the spherical HFB approximation.
Experimental data are taken from Ref.~\protect\cite{ref:mass}.
See Fig.~\protect\ref{fig:Sn_Dmass} for conventions.
\label{fig:Z8_S2n}}
\end{figure}

We present the two-neutron separation energies $S_{2n}$
for the O isotopes in Fig.~\ref{fig:Z8_S2n}.
The calculated values are compared with the experimental data.
We do not show $S_{2n}$ if the neutron chemical potential is positive.
Though not displayed in Fig.~\ref{fig:Z8_S2n},
$S_{2n}$ obtained from M3Y-P3 is close either to that from D1S or M3Y-P4.
Whereas $^{25-28}$O have experimentally been
established to be unbound~\cite{ref:O28},
most MF calculations so far have failed to reproduce this nature.
It would be noteworthy that
one of the present semi-realistic interactions, M3Y-P5,
correctly describes the location of the neutron drip line for oxygen
within the spherical HFB approximation;
$^{24}$O is the heaviest bound oxygen isotope.
The $^{25}$O nucleus has higher energy than $^{24}$O,
and in $^{26-28}$O the chemical potential becomes positive.
In contrast, $^{26}$O is bound in the HFB calculations
with M3Y-P3 and P4, as in the calculation with the D1S interaction.

\begin{figure}
\includegraphics[scale=1.0]{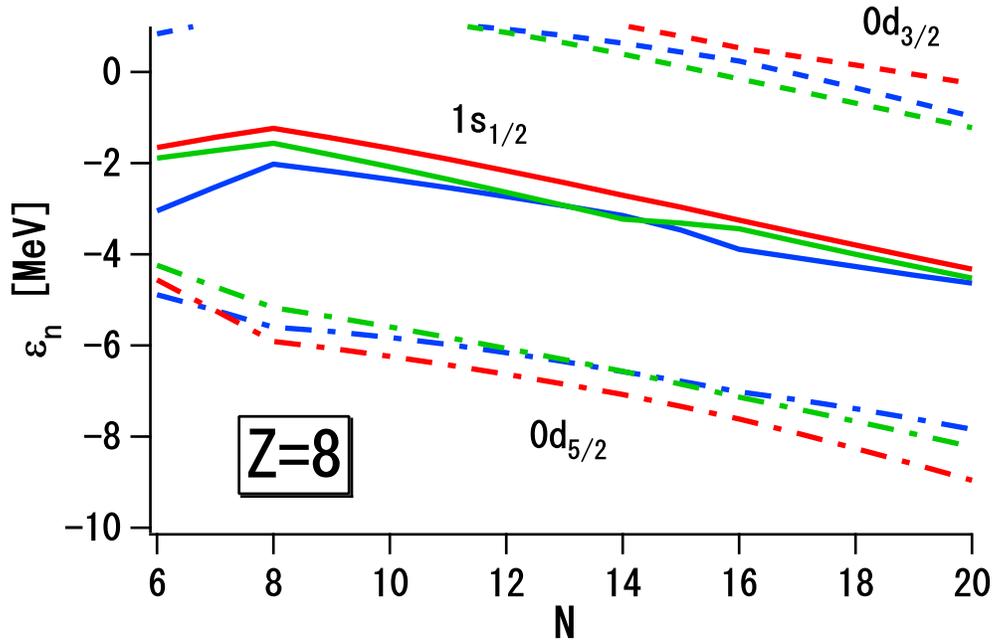}
\caption{HF single-particle energies in the O isotopes.
Blue, green and red lines represent the results
with the D1S, M3Y-P4 and M3Y-P5 interactions, respectively.
For each interaction, dot-dashed line is
for $\varepsilon_n(0d_{5/2})$,
solid line for $\varepsilon_n(1s_{1/2})$
and dashed line for $\varepsilon_n(0d_{3/2})$.
\label{fig:Z8_spe}}
\end{figure}

The reason why $^{26}$O is not bound with M3Y-P5
can be traced back to $\varepsilon_n(0d_{3/2})$,
the s.p. energy of $n0d_{3/2}$.
In Fig.~\ref{fig:Z8_spe}, the neutron s.p. energies
in the HF calculations are depicted.
We show the energies obtained from the bases of Eq.~(\ref{eq:basis-param})
even when the s.p. energy is positive,
not treating the boundary condition carefully.
We view that M3Y-P5 gives higher $\varepsilon_n(0d_{3/2})$
than the other interactions,
which originates from the slightly stronger $v^{(\mathrm{LS})}$
as well as from the relatively small $M^\ast_0$.
Note that we have fixed the enhancement factor for $v^{(\mathrm{LS})}$
in M3Y-P5 so as to reproduce the s.p. spectrum around $^{208}$Pb,
not adjusting accurately to,
\textit{e.g.}, the $\ell s$ splitting around $^{16}$O.
It is commented that $v^{(\mathrm{TN})}$ has small
but attractive contribution to $\varepsilon_n(0d_{3/2})$,
and therefore is irrelevant to the higher $\varepsilon_n(0d_{3/2})$
in the M3Y-P5 result.

\subsection{$Z=20$ and $28$ nuclei\label{subsec:drip-others}}

Location of the neutron drip line for the Ca and Ni nuclei
could be investigated by the currently constructed or designed
experimental facilities~\cite{ref:RIB07}.
We tabulate location of the neutron drip line
predicted by the spherical HFB calculations
with the M3Y-type and the D1S interactions,
in Table~\ref{tab:n-drip}.

\begin{table}
\begin{center}
\caption{Neutron numbers of the heaviest bound Ca and Ni nuclei
 predicted by the spherical HFB calculations
 with several interactions.
\label{tab:n-drip}}
\begin{tabular}{ccccc}
\hline\hline
Isotope &~~~D1S~~~&~M3Y-P3~&~M3Y-P4~&~M3Y-P5~\\ \hline
Ca & $44$ & $50$ & $48$ & $50$ \\
Ni & $58$ & $64$ & $62$ & $60$ \\
\hline\hline
\end{tabular}
\end{center}
\end{table}

\begin{figure}
\includegraphics[scale=1.0]{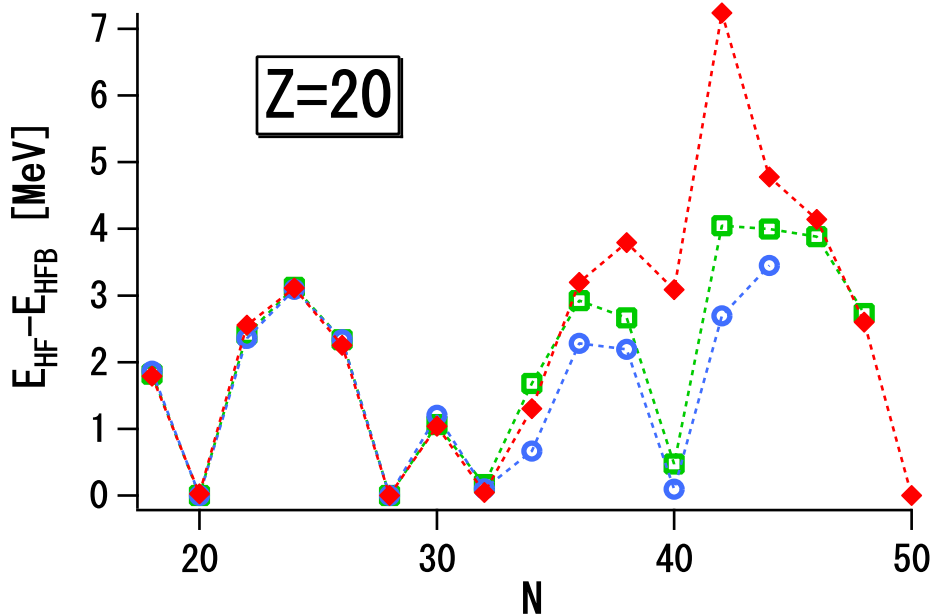}
\caption{Difference between the HF and HFB energies
for the Ca isotopes ($N=\mbox{even}$),
obtained from D1S, M3Y-P4 and P5.
See Fig.~\protect\ref{fig:Sn_Dmass} for conventions.
Dotted lines are drawn to guide eyes.
\label{fig:Z20_Epr}}
\end{figure}

If we use the D1S interaction,
the heaviest bound Ca nucleus is $^{64}$Ca,
because the neutron chemical potential is positive in $N\geq 45$.
The M3Y-P3 and P5 interactions predict that $^{70}$Ca is bound,
while $^{68}$Ca is the heaviest bound Ca isotope in the M3Y-P4 result.
We depict difference between the HF and the HFB energies
for the Ca nuclei in Fig.~\ref{fig:Z20_Epr},
which represents the pair correlation.
Though not shown, M3Y-P3 gives similar results to M3Y-P5.
While the pairing effects are in good agreement
among all of these interactions in $N\leq 32$,
the M3Y-type interactions give stronger pairing than D1S
in $N\geq 34$.

Both in the predicted position of the neutron drip line
and in the pair correlation in $N\geq 34$,
the s.p. energy of $n0g_{9/2}$ plays an important role.
The neutron drip line can extend up to $^{70}$Ca
if $\varepsilon_n(0g_{9/2})$ is sufficiently low.
At $^{60}$Ca we have $\varepsilon_n(0g_{9/2})=+0.73$, $-0.63$, $+0.23$
and $-0.65\,\mathrm{MeV}$ in the HF calculations
with D1S, M3Y-P3, P4 and P5, respectively,
well correlated to the location of the drip line.
The lower $\varepsilon_n(0g_{9/2})$ leads to
the smaller shell gap at $N=40$,
$\varepsilon_n(0g_{9/2})-\varepsilon_n(0f_{5/2})$,
which makes the pair excitation across $N=40$ easier.
The gap is $3.8$, $2.0$, $3.7$ and $1.7\,\mathrm{MeV}$
in D1S, M3Y-P3, P4 and P5.
It is noted that this quenching of the shell gap in M3Y-P3 and P5
comes from the relatively strong $v^{(\mathrm{LS})}$ to some degree,
but not from $v^{(\mathrm{TN})}$,
because $v^{(\mathrm{TN})}$ hardly contributes to the s.p. energies
in an $\ell s$-closed shell.

The vanishing difference between the HF and HFB energies
is often connected to the shell (or sub-shell) closure.
Figure~\ref{fig:Z20_Epr} indicates that, while $^{60}$Ca is stiff
against the pair excitation with D1S and M3Y-P4,
significant pair excitation occurs by M3Y-P5 (and by M3Y-P3),
because of the small shell gap at $N=40$.
On the contrary, $N=50$ is stiff against the pair excitation
in the present M3Y-P3 and P5 results, in which $^{70}$Ca is bound.
We have $\varepsilon_n(0g_{9/2})\approx -2\,\mathrm{MeV}$
at $^{70}$Ca,
and the gap between $n0g_{9/2}$ and the continuum
seems large enough for $N=50$ to keep the magic nature
against the pairing,
if we use the present semi-realistic pairing interaction.
Figure~\ref{fig:Z20_Epr} also suggests shell closure at $N=32$.
We shall return to this point in Sec.~\ref{sec:spe}.

\begin{figure}
\includegraphics[scale=1.0]{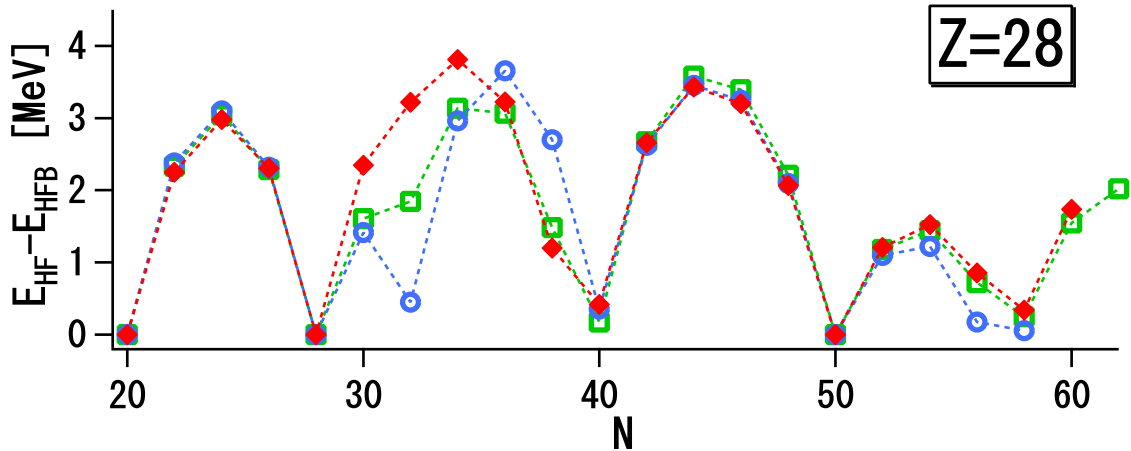}
\caption{Difference between the HF and HFB energies
for the Ni isotopes ($N=\mbox{even}$).
See Fig.~\protect\ref{fig:Sn_Dmass} for conventions.
\label{fig:Z28_Epr}}
\end{figure}

In Fig.~\ref{fig:Z28_Epr},
difference between the HF and HFB energies is shown
for the Ni isotopes.
For all the interactions,
the energy difference becomes vanishingly small at $N=20,28,40,50$
and $58$.
We view significant interaction-dependence in $28<N<40$.
In particular, the pair correlation is suppressed at $N=32$
with D1S, while no such effect is found with M3Y-P5.
It is noted that, with M3Y-P5, $^{68}$Ni seems almost doubly magic
as is consistent with experiments~\cite{ref:Ni68},
although $^{60}$Ca is not, as has been seen in Fig.~\ref{fig:Z20_Epr}.
For the neutron-rich Ni region,
energy sequence of the s.p. orbitals above $N=50$
is $1d_{5/2}$, $2s_{1/2}$, $1d_{3/2}$ and $0g_{7/2}$,
from the lower orbit to the higher.
The hindrance of the pair excitation at $^{86}$Ni
suggests magic or submagic nature of $N=58$
due to the gap between $n2s_{1/2}$ and $n1d_{3/2}$.
Unlike the M3Y-type interactions,
by D1S the pair excitation is hindered also at $N=56$.
The predicted neutron drip line appreciably depends
on $\varepsilon_n(1d_{3/2})$ and $\varepsilon_n(0g_{7/2})$.
The D1S interaction yields higher $\varepsilon_n(1d_{3/2})$
than the M3Y-type interactions,
which causes the drip line at $N=58$.
The low $\varepsilon_n(0g_{7/2})$ in M3Y-P3
induces pair excitation to $0g_{7/2}$,
leading to the binding up to $^{92}$Ni.

In the highly neutron-rich region,
the diffuseness of the nuclear surface becomes larger
than in the $\beta$-stable region.
Then the pair correlation could be relatively strong
if it has the surface-dominant nature.
However, it is not easy to argue precisely the extent of surface dominance
in the pairing from location of the drip line,
since it is obscured by influence of the shell structure.
In practice, when we use M3Y-P5 for the HF Hamiltonian
and D1S for the pair potential,
predicted location of the drip line for Ca is
the same as the result of the pure M3Y-P5 prediction.
For the Ni case, although $^{88}$Ni becomes unbound
when D1S is used for the pair potential,
the chemical potential is only $-0.07\,\mathrm{MeV}$
in the pure M3Y-P5 result.
It will be fair to say that the difference in the pairing channel
between the M3Y-type interactions and D1S
is not quite significant to location of the neutron drip line.
Dependence of the rms matter radii on the pairing interaction
is not apparent either,
as long as we work with the present M3Y-type or the D1S interactions,
whereas the radii near the drip line are sensitive
to the separation energies.

\section{Nucleus-dependence of single particle energies
\label{sec:spe}}

It has been pointed out that the shell structure,
particularly its nucleus-dependence
(sometimes called \textit{shell evolution}),
may be connected to characters
of effective interactions~\cite{ref:Nak03,ref:Vtn,ref:Vst}.
In recent studies role of the tensor force in the shell structure
has attracted great interest~\cite{ref:Vtn,ref:LBB07}.
It is known that
observed energies of one-particle states on top of a certain core
are appreciably disturbed by correlations beyond the MF regime.
However, it is presumable that those correlations do not vary
in a certain region of nuclei.
In the Sb and the $N=83$ nuclei,
s.p. energies of a few orbitals are extracted
from several fragmented states~\cite{ref:Sch04},
by averaging their energies weighted by the spectroscopic factors.
As a result, the averaged s.p. energies are shifted
from the lowest states with specific spin-parity
nearly by a constant, from nucleus to nucleus.
While data on the averaged energies are not available in many cases,
we proceed to investigate nucleus-dependence of the shell structure
by using the measured energies of the lowest states.

In this section we shall investigate nucleus-dependence of s.p. energies,
using the spherical HF or HFB calculations.
Results of several interactions are compared.
In the HF calculations for open-shell nuclei,
the HF Hamiltonian is obtained by folding the interaction
by the occupation numbers on each spherical orbital
up to the Fermi level.

\subsection{Neutron orbits and shell gap in $N=16$ nuclei
 \label{subsec:spe-N16}}

Nucleus-dependence of s.p. energies
could be relevant to the new magic numbers
in unstable nuclei~\cite{ref:Vst,ref:Vtn}.
We investigated the s.p. energies in the $N=16$ and $32$ nuclei
in Refs.~\cite{ref:Nak03,ref:Nak04},
and disclosed role of $v^{(\mathrm{C})}_\mathrm{OPEP}$
for the $N=16$ isotones,
using the M3Y-P2 interaction.
We shall reinvestigate nucleus-dependence of the s.p. energies
in these nuclei,
drawing attention also to the tensor force.

\begin{figure}
\includegraphics[scale=0.9]{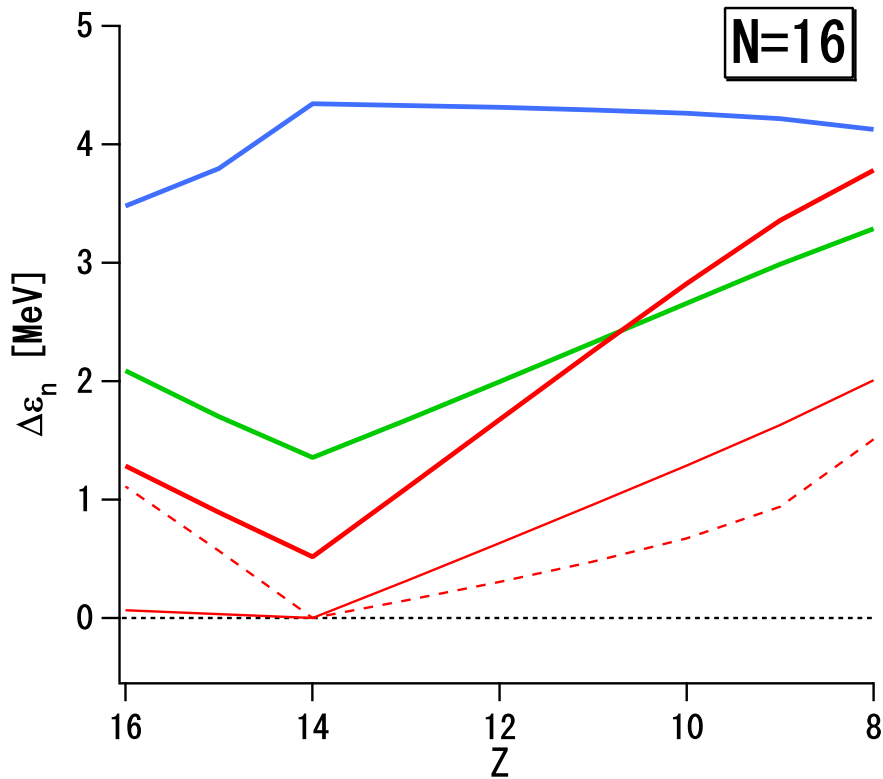}
\caption{${\mathit\Delta}\varepsilon_n=\varepsilon_n(0d_{3/2})
-\varepsilon_n(1s_{1/2})$ for the $N=16$ isotones.
Blue, green and red lines correspond to the results
with the D1S, M3Y-P4 and P5 interactions, respectively.
Thin red solid and dashed lines represent relative contributions
of $v^{(\mathrm{TN})}$ and $v^{(\mathrm{C})}_\mathrm{OPEP}$
in the M3Y-P5 result.
\label{fig:N16_dspe}}
\end{figure}

$Z$-dependence of the neutron s.p. energy $\varepsilon_n(0d_{3/2})$
relative to $\varepsilon_n(1s_{1/2})$ is appreciably affected
by effective interactions~\cite{ref:Nak02,ref:Nak03}.
Figure~\ref{fig:N16_dspe} depicts
${\mathit\Delta}\varepsilon_n=\varepsilon_n(0d_{3/2})
-\varepsilon_n(1s_{1/2})$ for varying $Z$
obtained from the HF calculations in the $N=16$ isotones.
Though not shown, ${\mathit\Delta}\varepsilon_n$ of M3Y-P3
resembles that of M3Y-P5.
To clarify role of $v^{(\mathrm{TN})}$
and $v^{(\mathrm{C})}_\mathrm{OPEP}$,
we also plot their contributions to ${\mathit\Delta}\varepsilon_n$
in the M3Y-P5 result,
which are calculated as $\sum_{j'}\langle N_{j'}\rangle\,
(2J+1)\langle jj'J|v|jj'J\rangle/(2j+1)(2j'+1)$
and shifted by the values at $N=14$.

The present semi-realistic interactions yield
increasing ${\mathit\Delta}\varepsilon_n$
as $Z$ goes from $14$ to $8$,
in contrast to the D1S interaction.
We view in Fig.~\ref{fig:N16_dspe} that
$v^{(\mathrm{C})}_\mathrm{OPEP}$~\cite{ref:Nak02}
and $v^{(\mathrm{TN})}$ produce this feature cooperatively.
This $Z$-dependence of the s.p. energies
could be relevant to the new magic number $N=16$
in the neutron-rich region~\cite{ref:N16}.
It has been confirmed that several popular Skyrme interactions
show similar behavior to D1S~\cite{ref:Nak02}.

\subsection{Neutron orbits and shell gap in $N=32$ nuclei
 \label{subsec:spe-N32}}

\begin{figure}
\includegraphics[scale=0.9]{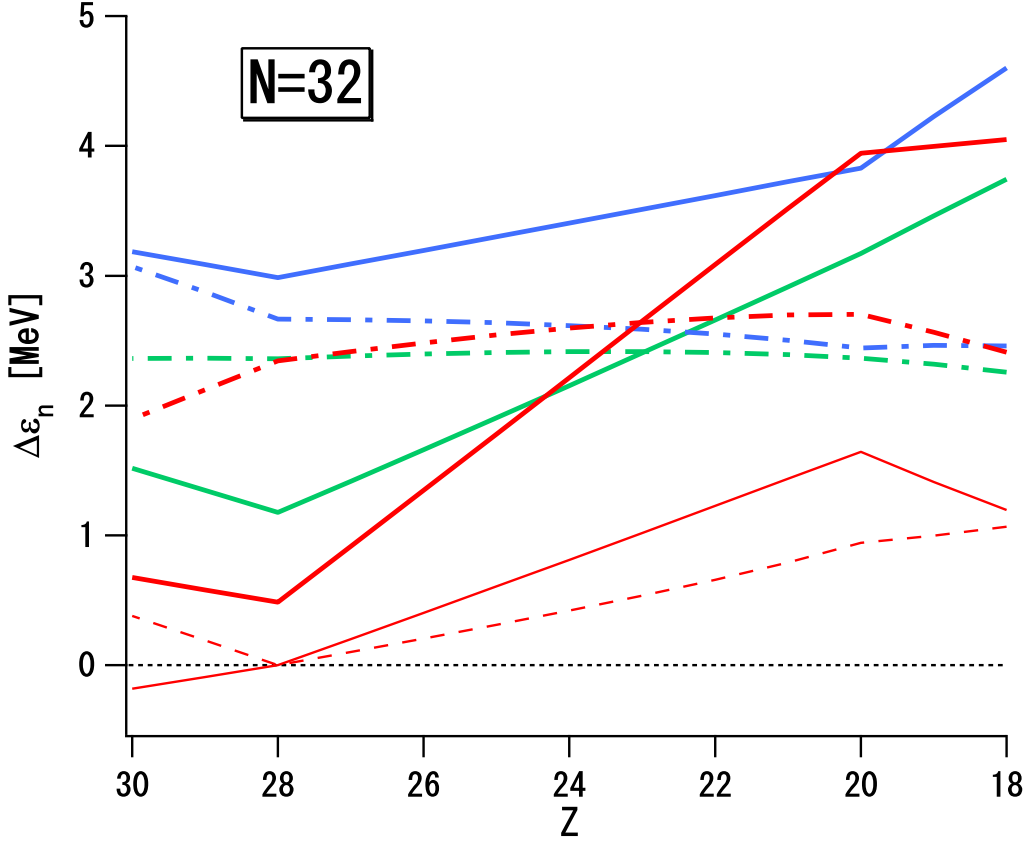}
\caption{${\mathit\Delta}\varepsilon_n(0f_{5/2})=\varepsilon_n(0f_{5/2})
-\varepsilon_n(1p_{3/2})$ (solid lines)
and ${\mathit\Delta}\varepsilon_n(1p_{1/2})=\varepsilon_n(1p_{1/2})
-\varepsilon_n(1p_{3/2})$ (dot-dashed lines) for the $N=32$ isotones.
See Fig.~\protect\ref{fig:N16_dspe} for conventions of colors.
Thin red solid and dashed lines represent relative contributions
of $v^{(\mathrm{TN})}$ and $v^{(\mathrm{C})}_\mathrm{OPEP}$
to ${\mathit\Delta}\varepsilon_n(0f_{5/2})$ in the M3Y-P5 result.
\label{fig:N32_dspe}}
\end{figure}

In Fig.~\ref{fig:N32_dspe},
$Z$-dependence of the neutron s.p. energies
relative to $\varepsilon_n(1p_{3/2})$,
${\mathit\Delta}\varepsilon_n(j)=\varepsilon_n(j)-\varepsilon_n(1p_{3/2})$,
is shown for the $N=32$ nuclei, by taking $j=0f_{5/2}$ and $1p_{1/2}$.
As in the preceding subsection, contributions of
$v^{(\mathrm{TN})}$ and $v^{(\mathrm{C})}_\mathrm{OPEP}$
to ${\mathit\Delta}\varepsilon_n(0f_{5/2})$ in the M3Y-P5 result
are also presented,
after shifting by the values at $Z=28$.
With the M3Y-P5 interaction we obtain strong $Z$-dependence
in ${\mathit\Delta}\varepsilon_n(0f_{5/2})$,
which could be relevant to the magicity of $N=32$
in the neutron-rich region~\cite{ref:Ca52_Ex2}.
Once again this $Z$-dependence originates in
$v^{(\mathrm{C})}_\mathrm{OPEP}$~\cite{ref:Nak04}
and $v^{(\mathrm{TN})}$.

This behavior of ${\mathit\Delta}\varepsilon_n(0f_{5/2})$
is reflected in the pair correlations shown in Figs.~\ref{fig:Z20_Epr}
and \ref{fig:Z28_Epr}.
The pairing effects are small at $N=32$
for all the interactions in the calcium case,
because ${\mathit\Delta}\varepsilon_n(0f_{5/2})$
as well as ${\mathit\Delta}\varepsilon_n(1p_{3/2})$
are greater than $2\,\mathrm{MeV}$.
Recall that this s.p. energy difference competes with the pairing gap,
whose typical value is estimated to be $\Delta\approx 12A^{-1/2}\approx
1.7\,\mathrm{MeV}$.
On the other hand, in the nickel case
the narrow ${\mathit\Delta}\varepsilon_n(0f_{5/2})$
leads to substantial pair excitation at $N=32$
for the M3Y-type interactions,
while such excitation is kept suppressed in the D1S result.

In the neutron-rich region of $Z\sim 20$,
there was a prediction that $N=34$ should be a magic number,
based on a shell model calculation~\cite{ref:HO02}.
The present MF calculations with the semi-realistic interactions
do not support this prediction.
While the $N=32$ shell gap is $2.7\,\mathrm{MeV}$ for $^{52}$Ca
in the HF calculation with M3Y-P5,
the $N=34$ gap is only $1.2\,\mathrm{MeV}$ for $^{54}$Ca.
The pair excitation across $N=34$ is sizable,
as has been viewed in Fig.~\ref{fig:Z20_Epr}.

\subsection{Proton orbits in $Z=50$ nuclei \label{subsec:spe-Z50}}

In recent studies,
nucleus-dependence of s.p. energies in the Sn isotopes
and in the $N=82$ isotones
has been disclosed from experiments~\cite{ref:Sch04}.
It has been pointed out that the tensor force seems to play
a crucial role in the $N$-dependence of the proton s.p. energies
in the Sn isotopes~\cite{ref:OMA06}.
We now have the M3Y-type interactions
with quite realistic tensor force (M3Y-P3 and P5)
and without tensor force (M3Y-P4),
both of which reproduce the properties of doubly magic nuclei
as well as the pairing properties to reasonable accuracy.
We apply the HFB calculations with these new interactions
to investigating the nucleus-dependence of the s.p. energies
in the Sn nuclei in this subsection,
and in the $N=50$ and $N=82$ nuclei in the subsequent subsection.

\begin{figure}
\includegraphics[scale=1.0]{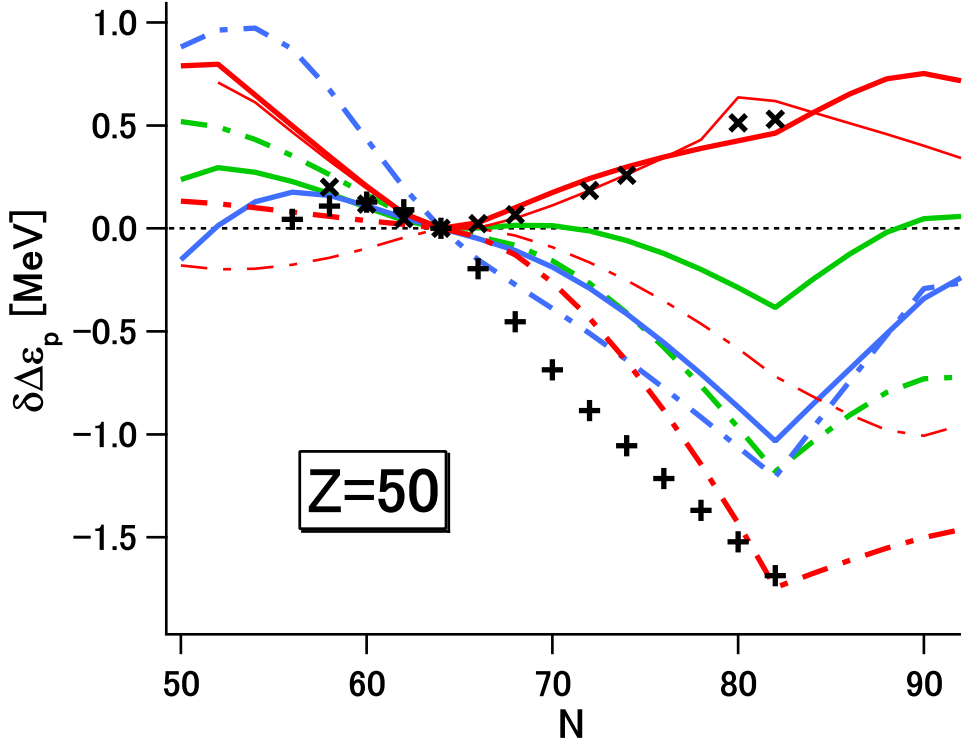}
\caption{$\delta{\mathit\Delta}\varepsilon_p(j)$ in the Sn isotopes
($N=\mbox{even}$), for $j=0g_{7/2}$ (dot-dashed lines)
and $0h_{11/2}$ (solid lines).
Blue, green and red lines represent the results of
D1S, M3Y-P4 and M3Y-P5, as before.
Pluses ($j=0g_{7/2}$) and crosses ($j=0h_{11/2}$) are experimental
values taken from the lowest states of the Sb nuclei~\protect\cite{ref:TI}.
Thin red lines are contributions of $v^{(\mathrm{TN})}$
in the s.p. levels of M3Y-P5.
\label{fig:Z50_dspe}}
\end{figure}

In the previous studies~\cite{ref:Sch04,ref:OMA06},
the relative proton s.p. energies
$\varepsilon_p(0h_{11/2})-\varepsilon_p(0g_{7/2})$
were the point of discussion.
We here consider the energies of these two orbits
relative to $1d_{5/2}$,
$\mathit{\Delta}\varepsilon_p(j)=\varepsilon_p(j)-\varepsilon_p(1d_{5/2})$
with $j=0g_{7/2}$ and $0h_{11/2}$.
Taking $\mathit{\Delta}\varepsilon_p(j)$ at $N=64$ to be a reference
and denoting it by $\mathit{\Delta}\varepsilon_p^0(j)$,
we plot $\delta\mathit{\Delta}\varepsilon_p(j)
=\mathit{\Delta}\varepsilon_p(j)-\mathit{\Delta}\varepsilon_p^0(j)$
in Fig.~\ref{fig:Z50_dspe}.
The values of the M3Y-type interactions (M3Y-P4 and P5)
are presented together with those of D1S and the experimental data.
For the data we use the energies of the lowest states in the Sb nuclei.

The semi-realistic M3Y-P5 interaction
reproduces $\delta\mathit{\Delta}\varepsilon_p(j)$
remarkably well.
In the $64\leq N\leq 82$ region,
the $N$-dependence of $\mathit{\Delta}\varepsilon_p(j)$
takes place due primarily to the occupation of $n0h_{11/2}$,
to which contribution of $v^{(\mathrm{TN})}$ is significant.
Though not shown, $\delta\mathit{\Delta}\varepsilon_p(j)$ with M3Y-P3
is close to the M3Y-P5 result.
With D1S, whereas $\delta\mathit{\Delta}\varepsilon_p(0g_{7/2})$
is in qualitative agreement with the data,
$\delta\mathit{\Delta}\varepsilon_p(0h_{11/2})$ is not,
because the tensor force is absent.
The same holds for M3Y-P4.
Although M3Y-P4 reproduces the tendency of the observed $N$-dependence
of $\varepsilon_p(0h_{11/2})-\varepsilon_p(0g_{7/2})$,
it gives wrong behavior
for $\delta\mathit{\Delta}\varepsilon_p(0h_{11/2})$;
\textit{i.e.}, $N$-dependence of $\varepsilon_p(0h_{11/2})$
relative to $\varepsilon_p(1d_{5/2})$.
It is also remarked that M3Y-P5 gives quite different behavior
of $\delta\mathit{\Delta}\varepsilon_p(j)$
from D1S and M3Y-P4 in $N<64$,
and that the currently available data favor the result of M3Y-P5.
We have confirmed that $v^{(\mathrm{C})}_\mathrm{OPEP}$
does not have important effects on the $N$-dependence
of $\mathit{\Delta}\varepsilon_p(j)$.

It is emphasized that the M3Y-P5 interaction can reproduce
the variation of the s.p. energy difference in the Sn isotopes
without destroying the shell structure of the doubly magic nuclei
shown in Sec.~\ref{sec:DMprop}.
This could be an advantage
of the present realistic tensor force.
In Ref.~\cite{ref:LBB07}, no such parameters were found
within the Skyrme density functional
including the zero-range tensor force.

\subsection{Neutron orbits in $N=50$ and $N=82$ nuclei
  \label{subsec:spe-N82}}

\begin{figure}
\includegraphics[scale=1.0]{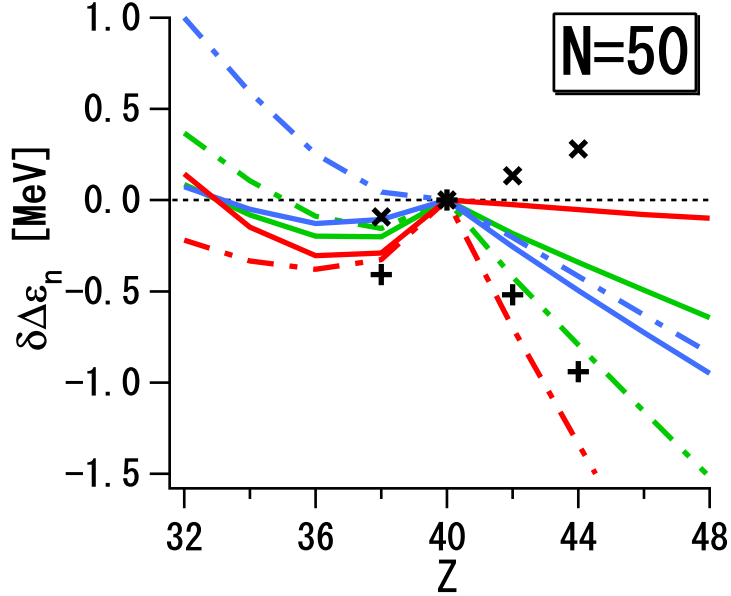}
\caption{$\delta{\mathit\Delta}\varepsilon_n(j)$ in the $N=50$ isotones
($Z=\mbox{even}$), for $j=0g_{7/2}$ and $0h_{11/2}$.
Conventions are the same as in Fig.~\protect\ref{fig:Z50_dspe},
except that $j$ represents the neutron orbits
and $Z=40$ is taken to be a reference.
Experimental values are taken from the lowest states
of the $N=51$ nuclei~\protect\cite{ref:TI}.
\label{fig:N50_dspe}}
\end{figure}

In the preceding subsection,
we have seen that the tensor force
affects $\varepsilon_p(0g_{7/2})$ and $\varepsilon_p(0h_{11/2})$
via the occupation of $n0h_{11/2}$.
This is accounted for by the attractive (repulsive) nature
of the tensor force between a neutron occupying a $j_>=\ell+1/2$ orbit
and a proton occupying $j'_<=\ell'-1/2$ ($j'_>=\ell'+1/2$)~\cite{ref:OMA06}.
The same mechanism is expected for $\varepsilon_n(0g_{7/2})$
and $\varepsilon_n(0h_{11/2})$ in the $N=50$ nuclei,
as $p0g_{9/2}$ is occupied.
We define $\mathit{\Delta}\varepsilon_n(j)=\varepsilon_n(j)
-\varepsilon_n(1d_{5/2})$,
and take its value at $Z=40$ to be $\mathit{\Delta}\varepsilon_n^0(j)$.
In Fig.~\ref{fig:N50_dspe}
$\delta\mathit{\Delta}\varepsilon_n(j)
=\mathit{\Delta}\varepsilon_n(j)-\mathit{\Delta}\varepsilon_n^0(j)$
is displayed for $j=0g_{7/2}$ and $0h_{11/2}$.
Although $\varepsilon_n(0g_{7/2})$ varies almost in parallel
to $\varepsilon_n(0h_{11/2})$ in $Z\geq 40$ for D1S,
notable $Z$-dependence arises for M3Y-P5,
in qualitative agreement with the observed s.p. levels.
The tensor force has significant contribution to this behavior
as in the $Z=50$ case.
More realistic than D1S in the central channels
but not having the tensor force,
M3Y-P4 yields $\delta\mathit{\Delta}\varepsilon_n(j)$ in-between
which is qualitatively good but quantitatively insufficient.

\begin{figure}
\includegraphics[scale=1.0]{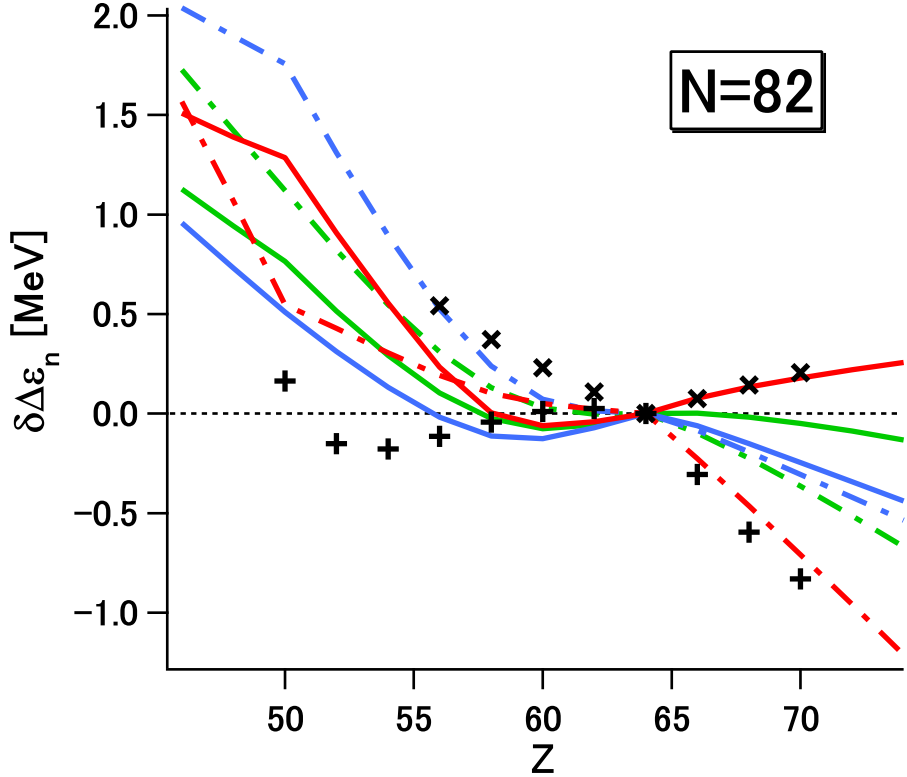}
\caption{$\delta{\mathit\Delta}\varepsilon_n(j)$
in the $N=82$ isotones ($Z=\mbox{even}$),
for $j=0h_{9/2}$ (dot-dashed lines)
and $0i_{13/2}$ (solid lines).
See Fig.~\protect\ref{fig:Z50_dspe} for conventions of colors.
Pluses ($j=0h_{9/2}$) and crosses ($j=0i_{13/2}$) are experimental
values taken from the lowest states
of the $N=83$ nuclei~\protect\cite{ref:TI}.
\label{fig:N82_dspe}}
\end{figure}

For the $N=82$ nuclei,
we consider $\mathit{\Delta}\varepsilon_n(j)
=\varepsilon_n(j)-\varepsilon_n(1f_{7/2})$,
from which $\delta\mathit{\Delta}\varepsilon_n(j)
=\mathit{\Delta}\varepsilon_n(j)
-\mathit{\Delta}\varepsilon_n^0(j)$ is obtained
by assuming the value at $Z=64$ as $\mathit{\Delta}\varepsilon_n^0(j)$.
Figure~\ref{fig:N82_dspe} shows $\delta\mathit{\Delta}\varepsilon_n(j)$
for $j=0h_{9/2}$ and $0i_{13/2}$.
The M3Y-P5 interaction well describes
$\delta\mathit{\Delta}\varepsilon_n(j)$ in $Z\geq 64$,
which is affected mainly by the occupation of $p0h_{11/2}$.
However, we cannot fully reproduce the tendency in $Z<64$,
which was argued in Ref.~\cite{ref:Sch04},
although the M3Y-P5 results of $\delta\mathit{\Delta}\varepsilon_n(j)$
are substantially better than those of D1S and M3Y-P4.
Further investigation will be necessary.

\section{Summary and outlook\label{sec:summary}}

We have developed semi-realistic effective interactions
to describe low energy phenomena of nuclei.
Starting from the M3Y interaction,
we add a density-dependent contact force
and modify several strength parameters in a phenomenological manner,
whereas maintaining the OPEP part in the central force.
We have obtained three new parameter-sets;
two of them (M3Y-P3 and P5) keep the tensor force
of the M3Y-Paris interaction,
and the other (M3Y-P4) has no tensor force.
Basic characters of the interactions are checked
by the Hartree-Fock calculations for the infinite nuclear matter,
and for the doubly magic nuclei.
The singlet-even channels of the interactions,
which are relevant to the pairing properties,
are fixed from the even-odd mass differences in the Sn isotopes,
by using the Hartree-Fock-Bogolyubov calculations.

We further implement the Hartree-Fock
and the Hartree-Fock-Bogolyubov calculations for spherical nuclei,
applying the new interactions.
Predicted shell structure depends on the effective interactions
to certain extent.
This may significantly affect location of the drip lines.
The new semi-realistic interaction M3Y-P5 correctly describes
the experimental consequence that the heaviest bound oxygen is $^{24}$O.
We have argued location of the neutron drip line
for the Ca and the Ni nuclei,
and its relevance to the shell structure.
Variation of the single-particle (s.p.) energies,
particularly contribution of the tensor force to it,
is a current topic.
It is suggested that the tensor force as well as the OPEP part
of the central force play a significant role
in the magic numbers $N=16$ and $32$ in the neutron-rich region.
It has been shown that the new semi-realistic interactions
including the tensor force, M3Y-P5 in particular,
describe the variation of the s.p. levels fairly well
in $Z=50$, $N=50$ and $N=82$ nuclei.
It is remarked that this interaction can reproduce
the variation of the s.p. energy difference in the Sn isotopes
without destroying the shell structure of the doubly magic nuclei.

It will be of interest to apply the semi-realistic interactions
to deformed nuclei, and to excited states
via the random-phase approximation (RPA).
Both projects are in progress
(for the latter, see Ref.~\cite{ref:Pb208}).

\begin{acknowledgments}
This work is financially supported
as Grant-in-Aid for Scientific Research (C), No.~19540262,
by the Ministry of Education, Culture, Sports, Science and Technology,
Japan.
Numerical calculations are performed on HITAC SR11000
at Institute of Media and Information Technology, Chiba University,
at Information Technology Center, University of Tokyo,
and at Information Initiative Center, Hokkaido University.
\end{acknowledgments}


\end{document}